\documentclass[aps,prd,nofootinbib,superscriptaddress,onecolumn,notitlepage,preprintnumbers]{revtex4-1}

\usepackage[english]{babel}
\usepackage[colorlinks=true,linkcolor=blue,citecolor=blue]{hyperref}

\usepackage{amsmath,graphicx,verbatim,epsfig}
\usepackage{amsmath,graphicx}
\usepackage{color}
\usepackage{datetime}

\usepackage{mathtools}

\usepackage{afterpage}

\newcommand{\sect}[1]{section~\protect\ref{#1}}

\newcommand{\gev}{\operatorname{GeV}}

\newcommand{\vek}[1] {\boldsymbol{#1}}

\newcommand{\y}{\vek{y}}
\newcommand{\ytilde}{\tilde{\vek{y}}}

\newcommand{\dd}{d}

\newcommand{\half}{{\textstyle\frac{1}{2}}}

\newcommand{\ms}{\mskip 1.5mu}

\allowdisplaybreaks[3]

\begin{document}

\preprint{NIKHEF 2014-042}


\title{Polarization effects in double open-charm production at LHCb}

\author{Miguel G. Echevarr\'ia}
\email{m.g.echevarria@nikhef.nl}
\affiliation{Nikhef, Science Park 105, 1098 XG Amsterdam, The Netherlands}
\affiliation{Department of Physics and Astronomy, VU
  University Amsterdam, De Boelelaan 1081, 1081 HV Amsterdam, The
  Netherlands}
  
\author{Tomas Kasemets}
\email{kasemets@nikhef.nl}
\affiliation{Nikhef, Science Park 105, 1098 XG Amsterdam, The Netherlands}
\affiliation{Department of Physics and Astronomy, VU
  University Amsterdam, De Boelelaan 1081, 1081 HV Amsterdam, The
  Netherlands}

\author{Piet J. Mulders}
\email{mulders@few.vu.nl}
\affiliation{Nikhef, Science Park 105, 1098 XG Amsterdam, The Netherlands}
\affiliation{Department of Physics and Astronomy, VU
  University Amsterdam, De Boelelaan 1081, 1081 HV Amsterdam, The
  Netherlands}

\author{Cristian Pisano}
\email{c.pisano@nikhef.nl}
\affiliation{Nikhef, Science Park 105, 1098 XG Amsterdam, The Netherlands}
\affiliation{Department of Physics and Astronomy, VU
  University Amsterdam, De Boelelaan 1081, 1081 HV Amsterdam, The
  Netherlands}
\affiliation{Department of Physics, University of Antwerp, Groenenborgerlaan 171, 2020 Antwerp, Belgium}



\begin{abstract}
Double open-charm production is one of the most promising channels to disentangle single from double parton scattering (DPS) and study different properties of DPS. Several studies of the DPS contributions have been made. A missing ingredient so far has been the study of polarization effects, arising from spin correlations between the two partons inside an unpolarized proton. We investigate the impact polarization has on the double open-charm cross section. We show that the longitudinally polarized gluons can give significant contributions to the cross section, but for most of the considered kinematic region only have a moderate effect on the shape. We compare our findings to the LHCb data in the $D^0D^0$ final state, identify observables where polarization does have an impact on the distribution of the final state particles, and suggest measurements which could lead to first experimental indications of, or limits on, polarization in DPS.
\end{abstract}

\maketitle

\section{Introduction}
\label{sec:intro}
Processes in hadron collisions where two partons from each hadron take part in separate partonic subprocesses, double parton scattering (DPS), contribute to several final states of interest at the LHC. DPS is a relevant background to precise Higgs boson coupling measurements and searches for new physics \cite{DelFabbro:1999tf,Hussein:2006xr,Bandurin:2010gn,Baer:2013yha}. The theory for DPS is still fragmentary, but major improvements have been made over the last couple of years moving towards a reliable description within perturbative QCD \cite{Manohar:2012jr,Diehl:2011tt,Blok:2010ge,Bansal:2014paa}. Despite this development there are still several important questions which have to be worked out.

For sufficiently inclusive cross sections DPS is formally a power suppressed contribution, but in certain regions of phase space double and single parton scattering contribute at the same power \cite{Diehl:2011yj}. Even for inclusive cross sections, DPS can in specific situations compete with single parton scattering - for example when the single parton scattering is suppressed by multiple small coupling constants. DPS is increasingly relevant with collider energy, and will hence be further enhanced when the LHC restarts to collide protons at larger center of mass energies. The reason is the rapid increase of the density of partons with energy and towards smaller $x$-fractions.

DPS signals have been measured at the LHC by both ATLAS \cite{Aad:2013bjm} and CMS \cite{Chatrchyan:2013xxa} in the $W$-boson plus dijet final state. Of particular interest for our present study is the LHCb measurements of double open-charm production \cite{Aaij:2012dz}, in final states such as $D^0D^0$. Among the most promising channels for a clean separation of double from single parton scattering are the production of two same sign $W$-bosons and double open charm quarks \cite{Luszczak:2011zp,Cazaroto:2013fua,Gaunt:2014rua,Gaunt:2010pi,vanHameren:2014ava,Maciula:2013kd,Golec-Biernat:2014nsa}. In fact, studies have shown that for double open-charm production in the kinematical region of the LHCb measurement, double parton scattering dominates over single parton scattering \cite{vanHameren:2014ava}.

DPS cross sections are factorized into two hard partonic subprocesses and two double parton distributions (DPDs). Only little is known about the size of the DPDs. They have been studied in a variety of quark models  \cite{Rinaldi:2014ddl,Rinaldi:2013vpa,Broniowski:2013xba,Chang:2012nw}, including correlations between the two partons inside the same proton. The correlations have generally been found to be sizable. An open question in double parton scattering is the effects of quantum-number correlations. These include correlations between the spins, colors, flavors and fermion numbers of the partons \cite{Kasemets:2014yna,Manohar:2012jr,Diehl:2011yj,Mekhfi:1988kj}. Upper limits on the DPDs describing quantum-number correlations have been derived \cite{Kasemets:2014yna, Diehl:2013mla}. For polarized DPDs these limits have further been shown to hold under radiative corrections from the leading-order double DGLAP evolution up to higher scales.

In particular the spin correlations (described by polarized DPDs) have direct relations to the directions of the final state particles, and thus have the potential to change both the sizes of the DPS cross sections and the distributions of the produced particles. For example, azimuthal modulations have been found for double vector boson production \cite{Kasemets:2012pr}. The effects of the quantum correlations on DPS cross sections have been calculated \cite{Kasemets:2012pr,Manohar:2012jr} but so far no numerical results at the cross section level have been obtained. Through studies of the scale evolution of the DPDs, limits on the degree of polarization and thereby its possible effect on DPS cross sections at different scales were set in \cite{Diehl:2013mla}. 

In this paper we examine the effect that polarization in DPS can have on the double $c\bar{c}$ production in kinematic regions resembling those of the LHCb $D^0D^0$ measurement \cite{Aaij:2012dz}. Several studies of this process already exist in the literature, but so far all have neglected the possibility of spin correlations. We demonstrate for the first time the quantitative impact of polarization on any DPS cross section.

The structure of the paper is as follows: In \sect{sec:gDPD} we discuss some basics of DPS with focus on polarization, introduce the different polarized and unpolarized double gluon distributions and discuss their scale evolution. In \sect{sec:xsec} we present the analytical results for the cross section calculation including all possible polarizations of two gluons in an unpolarized proton. In \sect{sec:model} we discuss the models for the DPDs which we use in order to obtain numerical results -- which we present and compare to LHCb data in \sect{sec:data}. We summarize our findings and discuss their implications in \sect{sec:concl}.

\section{Double gluon distributions}
\label{sec:gDPD}
\begin{figure}[htb]
  \centering
	\includegraphics[width=0.45\textwidth]{%
      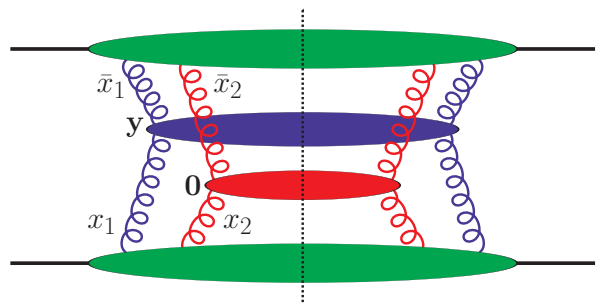}
  \caption{\label{fig:DPS} Gluon induced double parton scattering. The green fields represents the two DPDs while the blue and red fields represent the first and second hard interaction. $x_i$ ($\bar{x}_i$) are the longitudinal momentum fractions of the partons from the proton with momentum $p$ ($\bar{p}$). $\y$ and $\mathbf{0}$ are the transverse positions of the first and second hard interaction.}
\end{figure}

Under the assumption of factorization, as illustrated in figure \ref{fig:DPS}, the DPS cross section can be expressed schematically as
\begin{align}\label{eq:xsec}
\frac{d\sigma}{\prod_{i=1}^2 dx_i d\bar{x}_i }\Bigg|_{DPS} & = \frac{1}{C} \hat{\sigma}_1\hat{\sigma}_2 \int d^2\vek{y}\ F(x_1,x_2,\vek{y}) \bar{F}(\bar{x}_1,\bar{x}_2,\vek{y}) \, ,
\end{align}
where $\hat{\sigma}_i$ represents hard subprocess $i$ and $C$ is a combinatorial factor equal to two (one) if the final states of the two subprocesses are (not) identical. $F$ ($\bar{F}$) labels the double parton distribution of the proton with momentum $p$ ($\bar{p}$). The DPDs depend on the longitudinal momentum fractions of the two partons $x_i$ ($\bar{x}_i$) and their transverse separation $\vek{y}$. No complete proof for factorization in DPS exists, but several important ingredients have been established \cite{Manohar:2012jr,Diehl:2011yj}. The cross section expression \eqref{eq:xsec} is schematic as the labels for the different flavors, colors, fermion numbers and spins of the four partons are implicit. The possibility of interference between the two hard interactions, and correlations between the two partons inside each proton renders this structure significantly more complicated in DPS than for the case with only one hard interaction. Of particular interest for our purposes are the correlations between the spins of two gluons, and between the spins and the transverse separation, which lead to polarized gluon DPDs.

For the DPDs describing two gluons in an unpolarized right-moving proton we
write \cite{Diehl:2011yj}
\begin{align}
  \label{eq:dpds}
 F_{a_1a_2}(x_1,x_2,\vek{y})
 & = 2p^+ (x_1\ms p^+)^{-1}\, (x_2\ms p^+)^{-1}
        \int \frac{dz^-_1}{2\pi}\, \frac{dz^-_2}{2\pi}\, dy^-\;
           e^{i\ms ( x_1^{} z_1^- + x_2^{} z_2^-)\ms p^+}
\nonumber \\
 & \quad \times \left<p|\, \mathcal{O}_{a_2}(0,z_2)\, 
            \mathcal{O}_{a_1}(y,z_1) \,|p\right> \,.
\end{align}
We use light-cone coordinates $v^\pm = (v^0 \pm v^3) /\sqrt{2}$, bold font to denote the
transverse component $\vek{v} = (v^1, v^2)$ of any four-vector $v$ and $v_T=|\vek{v}|$.
The operators expressed in terms of the gluon field strength tensor reads
\begin{align}
\label{eq:gluon-ops}
  \mathcal{O}_{a_i}(y,z_i)
   &= \Pi_{a_i}^{jj'} \, G^{+j'}\bigl( y - \half z_i \bigr)\,
        G^{+j}\bigl( y + \half z_i \bigr)
   \Big|_{z_i^+ = y^+_{\phantom{i}} = 0,\; \vek{z}_i^{} = \vek{0}} \, ,
\end{align}
with projections
\begin{align}
  \label{eq:gluon-proj}
  \Pi_g^{jj'}  &= \delta^{jj'} \,,
&
  \Pi_{\Delta g}^{jj'} &= i\epsilon^{jj'} \,,
&
  [\Pi_{\delta g}^{kk'}]^{jj'} &= \tau^{jj'\!,kk'}
\end{align}
onto unpolarized gluons ($g$), longitudinally polarized gluons ($\Delta
g$) and linearly polarized gluons ($\delta g$). The tensor
\begin{align}
  \tau^{jj'\!,kk'} = \half \ms \bigl( \delta^{jk}\delta^{j'k'} 
     + \delta^{jk'}\delta^{j'k} - \delta^{jj'}\delta^{kk'} \bigr)
\end{align}
satisfies $\tau^{jj'\!,kk'} \tau^{kk'\!,\,ll'} = \tau^{jj'\!,\,ll'}$ and
is symmetric and traceless in each of the index pairs $(jj')$ and $(kk')$.

A decomposition of the
nonzero distributions for two gluons in terms of real-valued scalar
functions has been given in \cite{Diehl:2013mla}
\begin{align}
\label{eq:def-gg}
  F_{gg}(x_1,x_2,\y) & = f_{gg}(x_1,x_2,\y) \,,
  \nonumber\\
  F_{\Delta g \Delta g}(x_1,x_2,\y) & =
    f_{\Delta g \Delta g}(x_1,x_2,\y) \,,
  \nonumber\\
  F_{g \ms \delta g}^{jj'}(x_1,x_2,\y) & =
    \tau^{jj'\!,\y\y} M^2 f_{g \ms \delta g}(x_1,x_2,\y) \,,
  \nonumber\\
  F_{\delta g \ms g}^{jj'}(x_1,x_2,\y) & =
    \tau^{jj'\!,\y\y} M^2 f_{\delta g \ms g}(x_1,x_2,\y) \,,
  \nonumber\\
  F_{\delta g \delta g}^{jj',kk'}(x_1,x_2,\y) & =
    \half\ms \tau^{jj'\!,\,kk'} f_{\delta g \delta g}(x_1,x_2,\y) 
  + \bigl( \tau^{jj'\!,\y\ytilde} \tau^{kk'\!,\y\ytilde}
      - \tau^{jj'\!,\y\y} \tau^{kk'\!,\y\y} \bigr) \,
      M^4 f_{\delta g \delta g}^t(x_1,x_2,\y) \, ,
\end{align}
where $M$ is the proton mass and $\ytilde^j = \epsilon^{jj'} \y^{j'}$.  The notation where vectors $\y$ or
$\ytilde$ appear as an index of $\tau$ denote contraction, i.e.\
$\tau^{jj'\!,\y\y} = \tau^{jj'\!,kk'}\, \y^k \y^{k'}$ etc. The distributions of longitudinally polarized gluons carry open transverse indices $j,j',k,k'=\{1,2\}$ corresponding to the polarization vectors of the gluons which are contracted with the partonic cross sections.

The double parton distribution $f_{gg}$ represents the probability of finding two gluons with momentum fractions $x_1$ and $x_2$ at a transverse separation $\y$. The distribution of longitudinally polarized gluons $f_{\Delta g \Delta g}$ describe the difference in probability between finding the two gluons with their helicities aligned rather than anti-aligned, while linearly polarized gluons are described by helicity interference distributions, see for example \cite{Kasemets:2013nma,Diehl:2013mla} in the context of DPS. 

\subsection{Evolution of the double gluon distributions}
The scale evolution of the DPDs is governed by a generalization of the DGLAP evolution equations. Two versions exist in the literature: one homogenous equation describing two independent branchings of the two partons, and another including the splitting of a parent parton into the two partons which subsequently undergo hard scatterings \cite{Kirschner:1979im,Shelest:1982dg,Snigirev:2003cq,Gaunt:2009re,Ceccopieri:2010kg}. Which one is the correct one for describing DPS is still under debate \cite{Gaunt:2011xd,Gaunt:2012dd,Diehl:2011tt,Diehl:2011yj,%
Ryskin:2011kk,Ryskin:2012qx,Manohar:2012pe,Blok:2011bu,Blok:2013bpa,Snigirev:2014eua,Ceccopieri:2014ufa}. The contribution from the splitting term was investigated in \cite{Gaunt:2014rua} for double $c\bar{c}$ production and was seen to give a sizable contribution to the cross section, but also that the perturbative splitting preferred to take place early on -- and evolve as two separate branches for most of the evolution range. Including such a term in our study could naturally lead to an enhancement of the effect of the polarization and we will return to this discussion in section \ref{sec:concl}.  In the following we will make use of the homogeneous version, under the assumption that the physics of the single parton splitting contribution can be treated separately.

The evolution equation for the unpolarized double
gluon distribution then reads
\begin{align}
  \label{eq:evol-parton-1}
  \frac{\dd f_{gg}(x_1,x_2,\y; \mu)}{\dd \ln\mu^2}
    & = \frac{\alpha_s}{2\pi}\sum_{a=g,q,\bar{q}} \left[ P_{ga} \otimes_1 f_{ag}  + P_{ga} \otimes_2 f_{ga} \right] \,,
\end{align}
where
\begin{align}
\label{eq:otim}
    P_{ab}( \, .\, ) \otimes_1
       f_{bc}( \, .\, ,x_2,\y;\mu) 
    & = \int_{x_1}^{1-x_2} \frac{d u_1}{u_1}
        P_{ab}\left( \frac{x_1}{u_1} \right)
          f_{bc}(u_1,x_2,\y;\mu)\,,
          \nonumber\\
    P_{ab}( \, .\, ) \otimes_2
       f_{cb}( x_1, \, .\, ,\y;\mu) 
    & = \int_{x_2}^{1-x_1} \frac{d u_2}{u_2}
        P_{ab}\left( \frac{x_2}{u_2} \right)
          f_{bc}(x_1,u_2,\y;\mu)
\end{align}
are convolutions in the first and second argument of the DPDs with the leading-order
splitting kernels $P_{ab}$ known from DGLAP evolution of single-parton
distributions. Polarized DPDs follow equivalent evolution equations with the splitting kernels replaced by their polarized analogues. A more thorough discussion of the evolution of the polarized DPDs and expressions for all splitting kernels are given in \cite{Diehl:2013mla}. The evolution of the unpolarized gluon distribution leads to a violent increase at low momentum fractions, in particular at low scales where the QCD coupling constant is large. This is due to the $1/x$ behavior of the unpolarized splitting kernel in the limit where $x$ tends to zero. The splitting kernel for a longitudinally polarized gluon on the other hand approaches a constant in this limit, while the one for linearly polarized gluons goes as $x$. The polarized distributions therefore do not experience this rapid increase and evolution will suppress the relevance of polarized gluons -- in particular the linearly polarized ones. The rate at which this suppression takes effect leads to the expectation that at large scales (and not too large $x$) polarized gluons can be neglected in phenomenological calculations of DPS cross sections \cite{Diehl:2014vaa}. However, for double $c\bar{c}$ production the scales are low and there is only little room for evolution. This motivates the study of the effects of polarization in this process, and could, when confronted with experimental results lead to the first measurements of, or limits on, polarization effects in DPS.

\section{Double $c\bar{c}$ cross sections}
\label{sec:xsec}
We next present the analytic results of the cross section calculation, dividing the results into contributions from the different polarizations. The non-zero results come from gluons which are unpolarized, longitudinally polarized, mixed unpolarized-linearly polarized and purely linearly polarized.

\begin{figure}[htb]
  \centering
      	\includegraphics[width=0.2\textwidth]{%
      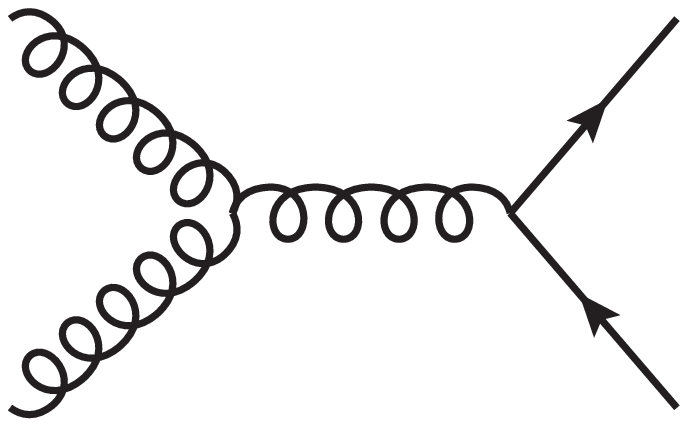}
      	\includegraphics[width=0.2\textwidth]{%
      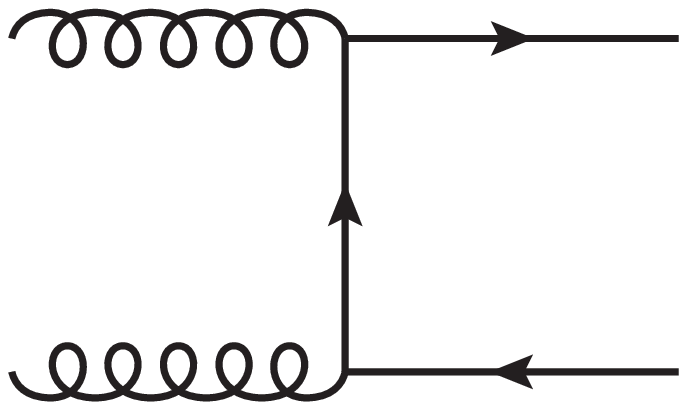}
      	\includegraphics[width=0.2\textwidth]{%
      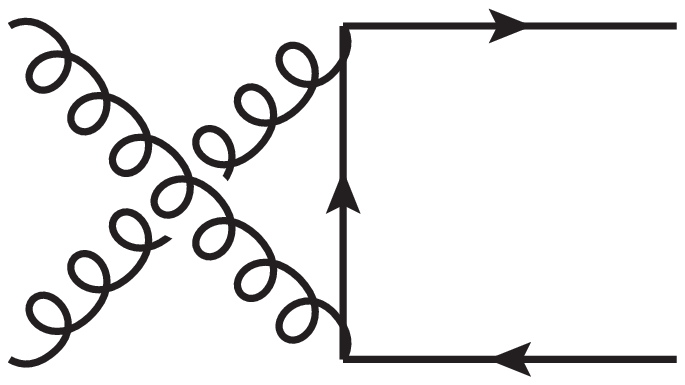}
  \caption{\label{fig:hard} Leading order Feynman diagrams contributing to each of the two partonic subprocesses. }
\end{figure}

Following experimental conventions, we present our results in the center-of-mass (CM) frame of the two protons, with $\hat{z}$-axis along the proton with momentum $p$ and $\hat{x}$-axis as pointing towards the centre of the LHC ring. With this choice of $\hat{x}$-axis, without reference to any direction defined by the process itself, any azimuthal dependence must show up as differences between the azimuthal angles describing the transverse directions of the final state particles. Double $c\bar{c}$ production in the kinematic region of interest is, to good approximation, initiated by gluons (see e.g. \cite{vanHameren:2014ava}). Therefore we limit ourselves to the partonic subprocesses of figure~\ref{fig:hard}, where the $c\bar{c}$ systems are produced by $s$-channel gluons or t-channel (u-chanel) charm quarks. 

The cross section contribution from unpolarized gluons reads
\begin{align}
	\frac{d\sigma_{(gg)(gg)}}{dy_1dy_2d^2\vek{p}_{1}d^2\vek{p}_2} & = \frac{1}{2} \left(8\pi^2\alpha_s^2\frac{N_c}{N_c^2-1}\right)^2 \int dx_1dx_2 
		 \prod_{i=1}^2\frac{1}{16\pi^2\hat{s}_i^2} \frac{2x_i\bar{x}_i}{2x_i-x_{Ti}e^{y_i}}
\nonumber\\
	 & \quad\times \frac{(1-z_1)^2+z_1^2-1/N_c^{2}}{(1-z_1)z_1} 
		 \left[(1-z_1^2)^2+z_1^2+4z_1(1-z_1) \Big(1-\frac{m^2}{m_{T1}^2}\Big)\frac{m^2}{m_{T1}^2}\right]
\nonumber\\
	 & \quad\times  \frac{ (1-z_2)^2+z_2^2-1/N_c^2 }{(1-z_2)z_2}
		 \left[(1-z_2^2)^2+z_2^2+4z_2(1-z_2) \left(1-\frac{m^2}{m_{T2}^2}\right)\frac{m^2}{m_{T2}^2}\right]
\nonumber\\
	& \quad\times \int d^2\y f_{gg}(x_1,x_2,\y)\bar{f}_{gg}(\bar{x}_1,\bar{x}_2,\y) \, ,
\end{align}
where $y_i$ and $\vek{p}_i$ are the rapidity and transverse momentum of the charm quark produced in interaction $i=1,2$. The variables in the cross section are given by
\begin{align}
z_i & = \frac{m^2-\hat{t}_i}{\hat{s}_i} = -\frac{x_{Ti}}{2\bar{x}_{i}}e^{-y_{i}}, & 
\bar{x}_i &=\frac{x_ix_{Ti}e^{-y_i}}{2x_i-x_{Ti}e^{y_i}}, &
x_{Ti} & = \frac{2m_{Ti}}{\sqrt{s}}, &
m_{Ti} = \sqrt{\vek{p}_{i}^2 + m^2}.
\end{align}
$\hat{s}_i$ and $\hat{t}_i$ are the usual Mandelstam variables of the partonic cross section $i$. $s$ is the center of mass energy of the proton collision, $m$ is the charm mass and $y_i$ is the rapidity of the charm quark from interaction $i$.

For gluons with longitudinal polarization the cross section is
\begin{align} \label{eq:xseclong}
	\frac{d\sigma_{(\Delta g \Delta g)(\Delta g \Delta g)}}{dy_1dy_2d^2\vek{p}_1d^2\vek{p}_2} & = \frac{1}{2}\left(8\pi^2\alpha_s^2\frac{N_c}{N_c^2-1}\right)^2\int dx_1dx_2 
		 \prod_{i=1}^2\frac{1}{16\pi^2\hat{s}_i^2} \frac{2x_i\bar{x}_i}{2x_i-x_{Ti}e^{y_i}}
\nonumber\\
	 & \quad\times  \frac{(1-z_1)^2+z_1^2-1/N_c^{2}}{(1-z_1)z_1} \left( 1 - \frac{2m^2}{m_{T1}^2} \right) \left[(1-z_1)^2+z_1^2 \right]
\nonumber\\
	 & \quad\times  \frac{(1-z_2)^2+z_2^2-1/N_c^{2}}{(1-z_2)z_2} \left( 1 - \frac{2m^2}{m_{T2}^2} \right) \left[(1-z_2)^2+z_2^2 \right]
\nonumber\\
	& \quad\times \int d^2\y f_{\Delta g \Delta g}(x_1,x_2,\y)\bar{f}_{\Delta g \Delta g}(\bar{x}_1,\bar{x}_2,\y).
\end{align}
Worth noticing is that differences in the partonic cross section between longitudinally and unpolarized gluons are suppressed by $m^2/m_{Ti}^2$. For transverse momenta of the outgoing charm quarks above a few GeV, this suppression is strong and already at $p_{Ti}=3\gev$ we have $m^2/m_{Ti}^2 = 0.16$. This tells us that differences in the distributions of the final state charm quarks produced at large $p_T$ between longitudinally polarized and unpolarized gluons will to good approximation originate in differences between $f_{gg}$ and $f_{\Delta g \Delta g}$.

Unpolarized mixed with linearly polarized gluons gives the cross section
\begin{align}
	\frac{d\sigma_{(\delta g g)(g \delta g)}}{dy_1dy_2d^2\vek{p}_1d^2\vek{p}_2} & = \left(8\pi^2\alpha_s^2\frac{N_c}{N_c^2-1}\right)^2\int dx_1dx_2 
		  \prod_{i=1}^2\frac{1}{16\pi^2\hat{s}_i^2} \frac{2x_i\bar{x}_i}{2x_i-x_{Ti}e^{y_i}}
\nonumber\\
	 & \quad\times \left ((1-z_1)^2+z_1^2-1/N_c^{2} \right) \frac{m^2}{m_{T1}^2} \left( 1 - \frac{m^2}{m_{T1}^2} \right) 
\nonumber\\
	 & \quad\times \left(  (1-z_2)^2+z_2^2-1/N_c^{2} \right) \frac{m^2}{m_{T2}^2} \left( 1 - \frac{m^2}{m_{T2}^2} \right)
\nonumber\\
	& \quad\times \cos (2\Delta \phi) \int d^2\y \; \y^4M^4f_{\delta g g}(x_1,x_2,\y)\bar{f}_{ g \delta g}(\bar{x}_1,\bar{x}_2,\y),
\end{align}
and the same result for the case when the linear and unpolarized gluons are interchanged, i.e. with $g \leftrightarrow \delta g$. $\Delta \phi = \phi_1-\phi_2$, where $\phi_i$ is the azimuthal angle of the outgoing $c$-quark from hard interaction $i$. The $\cos (2\Delta\phi)$ dependence is an effect of the difference in helicity between the amplitude and conjugate amplitude for the linearly polarized gluons. This term gives rise to the same kind of modulation in the azimuthal angle as observed by LHCb in the $D^0D^0$ final state \cite{Aaij:2012dz}. However, in our leading order cross section, the whole contribution for the mixed linear-unpolarized gluons is suppressed by $m^2/m_{Ti}^2$ for each of the two hard subprocesses. This suppression gives low analyzing power, and indicates already at this level that the contribution should be small. The suppression arises in the terms where there is an helicity flip in the hard cross section. For zero quark masses these terms in the partonic cross section tend to zero. The nonzero charm mass allows for a nonzero result of the mixed un- and linearly-polarized gluons, but only at the price of the suppression factor. This has previously been discussed in the context of heavy quark production with transverse momentum dependent parton distributions \cite{Pisano:2013cya}. It is however interesting to note that the suppression could be lifted if the gluons are given a transverse momentum, for example by radiating off a gluon. The next-to-leading-order (NLO) correction to the cross section is expected to be large \cite{vanHameren:2014ava}, and a large NLO contribution in combination with a lifting of the suppression have the potential to result in a large enhancement of this contribution. We will return to this point in the discussion of \sect{sec:concl}.

The cross section for gluons with linear polarization is
\begin{align}
	\frac{d\sigma_{(\delta g \delta g)(\delta g \delta g)}}{dy_1dy_2d^2\vek{p}_1d^2\vek{p}_2} & = \frac{1}{4}\left(8\pi^2\alpha_s^2\frac{N_c}{N_c^2-1}\right)^2\int dx_1dx_2 
		\prod_{i=1}^2\frac{1}{16\pi^2\hat{s}_i^2} \frac{2x_i\bar{x}_i}{2x_i-x_{Ti}e^{y_i}}
\nonumber\\
	 & \quad\times \left ((1-z_1)^2+z_1^2-1/N_c^{2} \right) \left(  (1-z_2)^2+z_2^2-1/N_c^{2} \right)
\nonumber\\
	 & \quad\times  \left[ \left(1-\frac{m^2}{m_{T1}^2} \right)^2\left(1-\frac{m^2}{m_{T2}^2}\right)^2 \cos (4\Delta \phi) +\frac{m^8}{m_{T1}^4m_{T2}^4} \right]
\nonumber\\
	& \quad\times \int d^2\y f_{\delta g \delta g}(x_1,x_2,\y)\bar{f}_{\delta g \delta g}(\bar{x}_1,\bar{x}_2,\y).
\end{align}
The pure linearly polarized contribution to the cross section has a $\cos(4\Delta\phi)$ dependent part and a $\Delta\phi$ independent part. The $\Delta\phi$ independent term comes with a double helicity flip in the two hard cross sections and is heavily suppressed. Notice that the $f^t_{\delta g \delta g}$ term in \eqref{eq:def-gg} does not contribute. This is because it results in a dependence on the angle between the directions of the outgoing charm quarks and the direction $\y$ between the two hard subprocesses, which vanishes upon integration over $\y$.

\section{Simple model for DPDs}
\label{sec:model}
In order to obtain numerical results we need an initial ansatz for the DPDs at some low starting scale. We decompose the unpolarized DPDs into two single parton distributions and a $\y$ dependent function assumed to be universal and $x_i$ independent 
\begin{align}\label{eq:dpd-factorized}
f_{gg}(x_1,x_2,\y;Q_0) = f_g(x_1,Q_0)f_g(x_2;Q_0)G(\y).
\end{align}
This is an ansatz commonly used for DPS phenomenology but its validity is questionable and in some kinematic regions wrong. The easiest way to see this is in the region of large $x_i$. Momentum conservations forces $x_1+x_2 \leq 1$ on the left side of \eqref{eq:dpd-factorized}, but the right hand side does, as it stands, not respect this constraint and gives nonzero values as long as both momentum fractions individually are below 1. A way to reinstate this limit is to multiply the ansatz with the factor $(1-x_1-x_2)$ to some positive power, however for the charm production the contribution of the large $x_i$ region is negligible and we apply a strict cutoff at the kinematic limit. Despite its limitations, the ansatz provides a useful starting point for DPS studies and we use it as input for the unpolarized DPDs at some low starting scale. These input distributions will then be evolved to higher scales with the double DGLAP equations. The numerical results will only be given in terms of ratios of cross sections, in which the $\y$ dependence cancels. For unpolarized distributions, the difference between separately evolving the two parton distribution functions (PDFs) or evolving the DPD with the factorized initial ansatz is small, except in the large $x_i$ region \cite{Diehl:2014vaa}. 

For polarized distributions, which describe the correlation between the spin of the two partons, it does not make sense to decompose it into polarized single parton distributions -- which describe the correlation between the spin of one parton and the spin of the proton. Instead we use the positivity bounds in \cite{Diehl:2013mla} to set upper limits on the sizes of the polarized distributions in terms of the unpolarized. We are interested in examining the maximal effects possible from the different polarizations and therefore saturate the bounds for each polarized DPD independently. This results in the relations 
\begin{align}
f_{\Delta g \Delta g} (x_1,x_2,\y;Q_0) & = f_{gg}(x_1,x_2,\y;Q_0) ,\nonumber\\
f_{g \delta g}  (x_1,x_2,\y;Q_0)& = \y^2M^2 f_{gg} (x_1,x_2,\y;Q_0), \nonumber\\
f_{\delta g \delta g}  (x_1,x_2,\y;Q_0)& = \y^4M^4 f_{gg} (x_1,x_2,\y;Q_0)
\end{align}
at the initial scale $Q_0$. If the bounds are fulfilled at an initial scale they will remain valid at all larger scales, but typically be violated at lower scales. We therefore choose to saturate the bounds at some low $Q_0$, and use the double DGLAP evolution \eqref{eq:evol-parton-1} with polarized splitting kernels to obtain the polarized DPDs at higher scales.

A larger $Q_0$ gives less room for evolution and therefore less suppression of the polarized contribution to the cross section. $Q_0$ should be chosen such that one is in a regime where perturbative QCD is expected to give sensible results. For the usual PDFs, the starting scale is often chosen somewhere around 1-2 GeV, and several of the leading order distributions go negative when evolved backwards to scales below 1 GeV. Another issue is the large uncertainty even for the distribution of a single unpolarized gluon at small scales and momentum fractions. The smaller values we take for the initial scale the larger this uncertainty is, and we would like to stay at a scale where we can compare between different sets of single parton distributions. We use this as a guidance in choosing starting scales at which to saturate our polarized bounds, and conclude that a choice somewhere between 1 and 2 GeV is reasonable. We investigate the impact of this choice by varying the input scale between the two values. 

For input PDFs we will use the leading-order GJR distributions \cite{Gluck:2007ck}. At these scales there are still large differences between different PDF sets, and we have investigated how they influence our results by switching to the MSTW2008lo distributions \cite{Martin:2009iq}. There are clear differences, especially with $Q_0=1\gev$, but the differences are smaller than those obtained by changing from $Q_0=1\gev$ to $Q_0=2\gev$. As a rule of thumb, the MSTW distributions give smaller polarization than the GJR distributions -- see \cite{Diehl:2014vaa} for more details on the effects of changing between different sets of PDFs. We use the values of $\alpha_s$ and the charm mass $m$ used in the PDF sets for concistency.

In addition, we examine the effect of changing our modeling of the polarized DPDs. Instead of taking the maximal allowed polarization, we can create a model built on ratios of splitting kernels describing the branching of a parent parton into the two gluons which subsequently undergo hard scatterings. For the longitudinally polarized DPD this results in
\begin{align}
f_{\Delta g \Delta g} = \frac{T_{g\rightarrow \Delta g \Delta g}}{T_{g\rightarrow g g}} f_{gg} = \frac{zz'(2-zz')}{z^2+z'^2+z^2 z'2} f_{gg}
\end{align}
where $z=x_1/(x_1+x_2)$ and $z'=1-z$. This model has been described in more detail in \cite{Diehl:2014vaa}, which gives a complete list of expressions for the different DPDs. We will only display results obtained in this model for longitudinally polarized gluons.

\section{Numerical results and comparison with data}
\label{sec:data}

We next turn to the numerical evaluation of the cross section in the kinematic regions probed by the double open-charm measurement by the LHCb Collaboration \cite{Aaij:2012dz}. The two charm quarks are required to have a transverse momentum in the region $3 \leq p_{Ti} \leq 12 \gev$ and rapidities in the range $2 \leq y_i \leq 4$, at $\sqrt{s}=7$~TeV.  The phase-space of the two anticharm quarks is integrated over, without experimental cuts, since they remain undetected.

The DPD evolution equations are evaluated by the code described in \cite{Gaunt:2009re}, which has been modified to suit our purposes as described in \cite{Diehl:2014vaa}. The main modifications are the use of the homogeneous evolution equations and the incorporation of the polarized splitting kernels for the evolution of the polarized DPDs, listed in appendix A of \cite{Diehl:2013mla}. We generated gridfiles for the DPDs in the range $10^{-6} \leq x_i \leq 1$ with 240 gridpoints in each direction, and 60 points in $\ln \mu^2$ in the range $Q^2_0 < \mu^2 < 2\times10^6\gev$. The phase-space integrations were performed numerically.

Care must be taken when comparing the data to the results of our calculation. While the calculation produces two pairs of charm-anticharm quarks, out of which only the two charm quarks are measured, the data is for $D^0D^0$. Simply interpreting the variables of the charm quarks as those of the final state mesons neglects the effects from hadronization/fragmentation. The assumption that the direction of the charm quark is approximately equal to that of the $D^0$ is commonly made \cite{Maciula:2013kd}. For the effect on the absolute size of the transverse momenta, the approximation is less accurate, but charm fragmentation functions typically peak around rather large $z$ values \cite{Peterson:1982ak,Braaten:1994bz}. However this approximation on the normalized cross section is not likely to change the spectrum at the level of precision we are interested in here. Normalizing the results to the total cross section cancels the effects on the absolute size, such as the branching ratio of $c\rightarrow D^0$. Our primary purpose is not to make exact predictions for the $D^0D^0$ cross section, but rather to examine the effects that polarization has on double charm production.
\begin{figure}[htb]
  \centering
	\includegraphics[width=0.4\textwidth]{%
      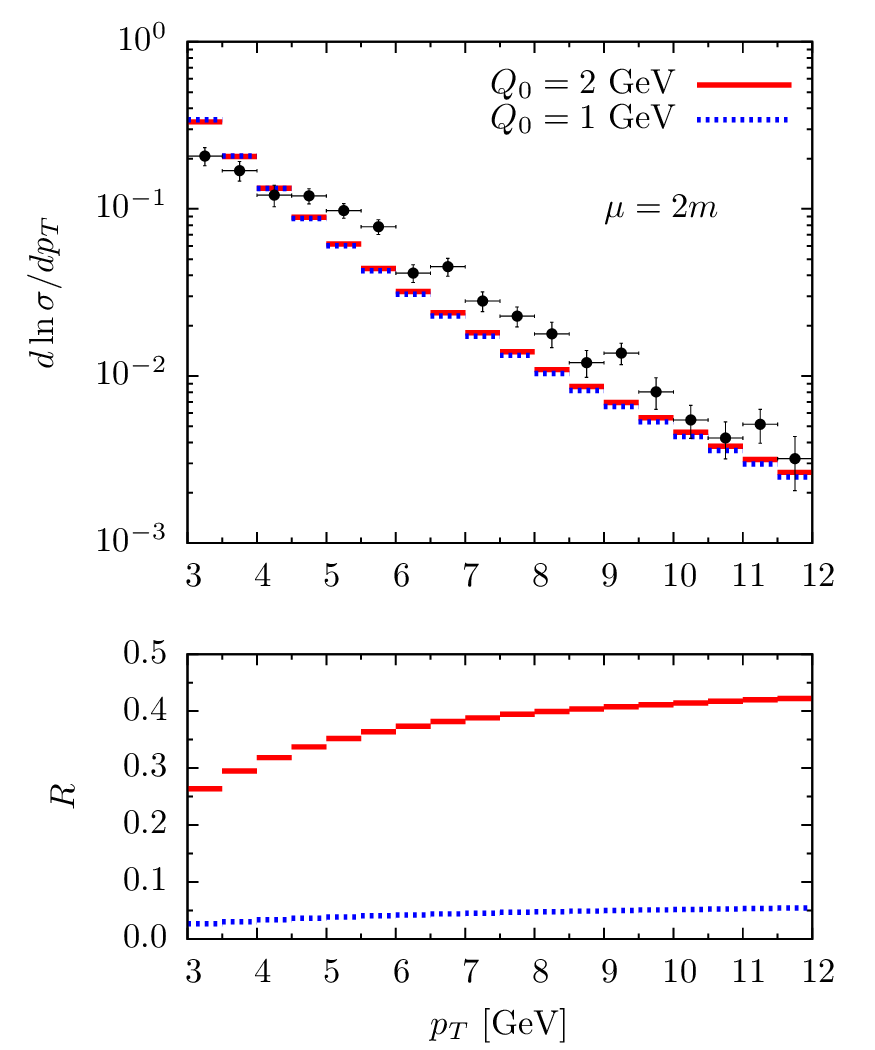}
 	\includegraphics[width=0.4\textwidth]{%
      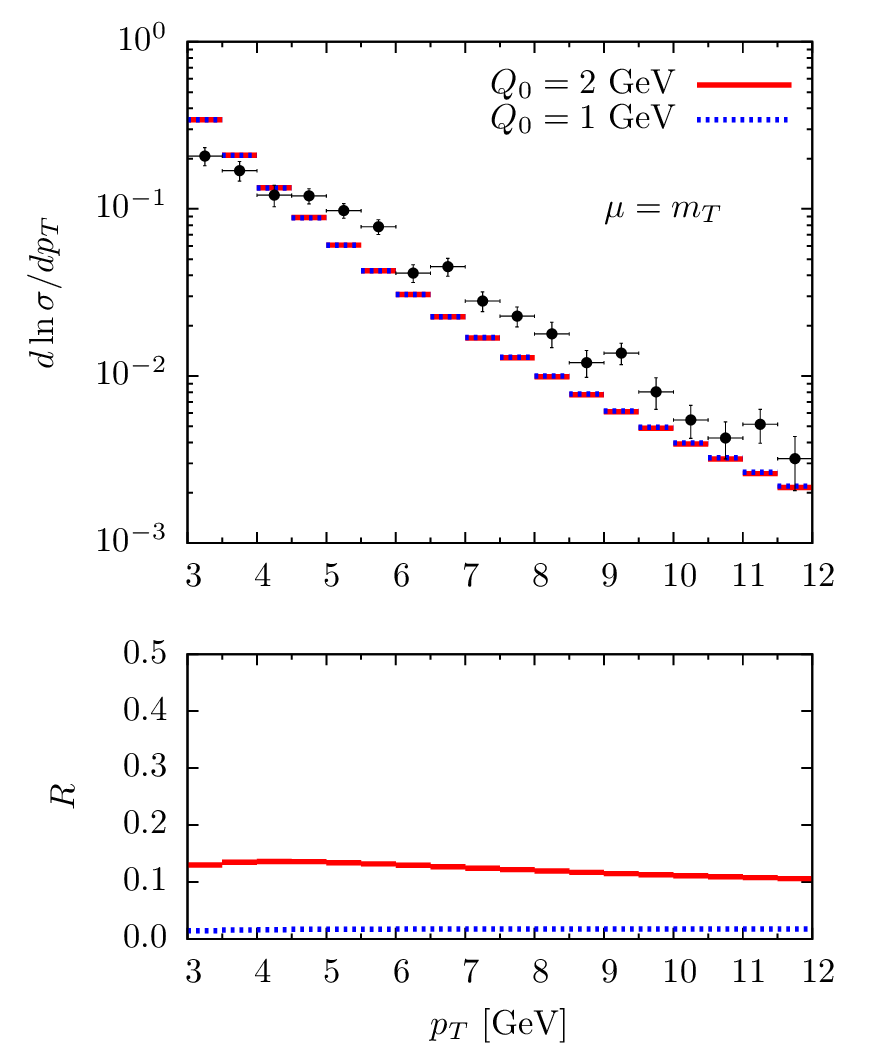}
  \caption{\label{fig:pT} Normalized cross section vs the transverse momentum of one of the charm quarks at $\mu=2m$ (left) and $\mu=m_T$ (right). Overlaid are the LHCb $D^0D^0$ data \cite{Aaij:2012dz}. The lower panels show the relative size of the polarized contribution. }
\end{figure}

Figure \ref{fig:pT} shows the dependence of the normalized cross section on the transverse momentum of one of the two charm quarks, $p_T$, as well as the ratio of the polarized over unpolarized contribution. The left panel shows the results with the scale choice of $\mu = 2m$ and the right with $\mu=m_T$, where 
\begin{equation}
m_T=\frac{m_{T1}+m_{T2}}{2}
\end{equation}
is the average transverse mass of the two charm quarks. For both cases, we make visible the dependence of the result on the choice of input scale by displaying the results for $Q_0=1\gev$ and $Q_0=2\gev$, as discussed in section \ref{sec:model}. The two lower panels show the relative size of the polarized contribution compared to the unpolarized. The cross section result in figure \ref{fig:pT} reproduces the data reasonably well. The shape of the cross section only has a tiny dependence on the choice of $Q_0$, whilst the contribution from the polarized distributions changes with $Q_0$. Likewise, there is little difference in the shape of the cross section with the two scale choices, but the polarized contribution is larger for $\mu=2m$ than for  $\mu=m_T$. This is expected since the latter choice allows for a larger evolution range and thus a stronger enhancement of the unpolarized over the polarized DPDs. With $\mu=m_T$ the suppression due to evolution also increases with the $p_T$, since increasing $p_T$ increases $m_T$, counteracting the enhancement of the polarized contribution from the partonic cross sections. The relative size of the polarized contribution does have a small dependence on $p_T$ with $\mu=2m$, but is rather flat for $\mu=m_T$. 
\begin{figure}[htb]
  \centering
	\includegraphics[width=0.4\textwidth]{%
      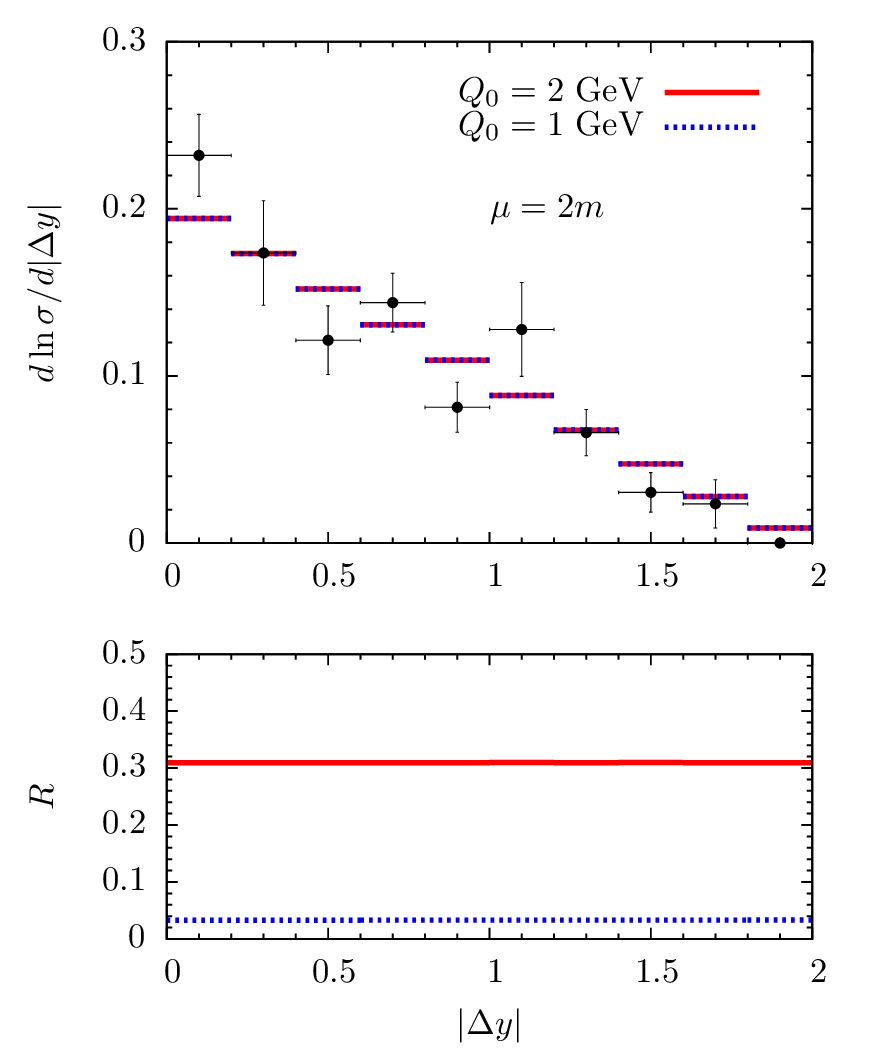}
 	\includegraphics[width=0.4\textwidth]{%
      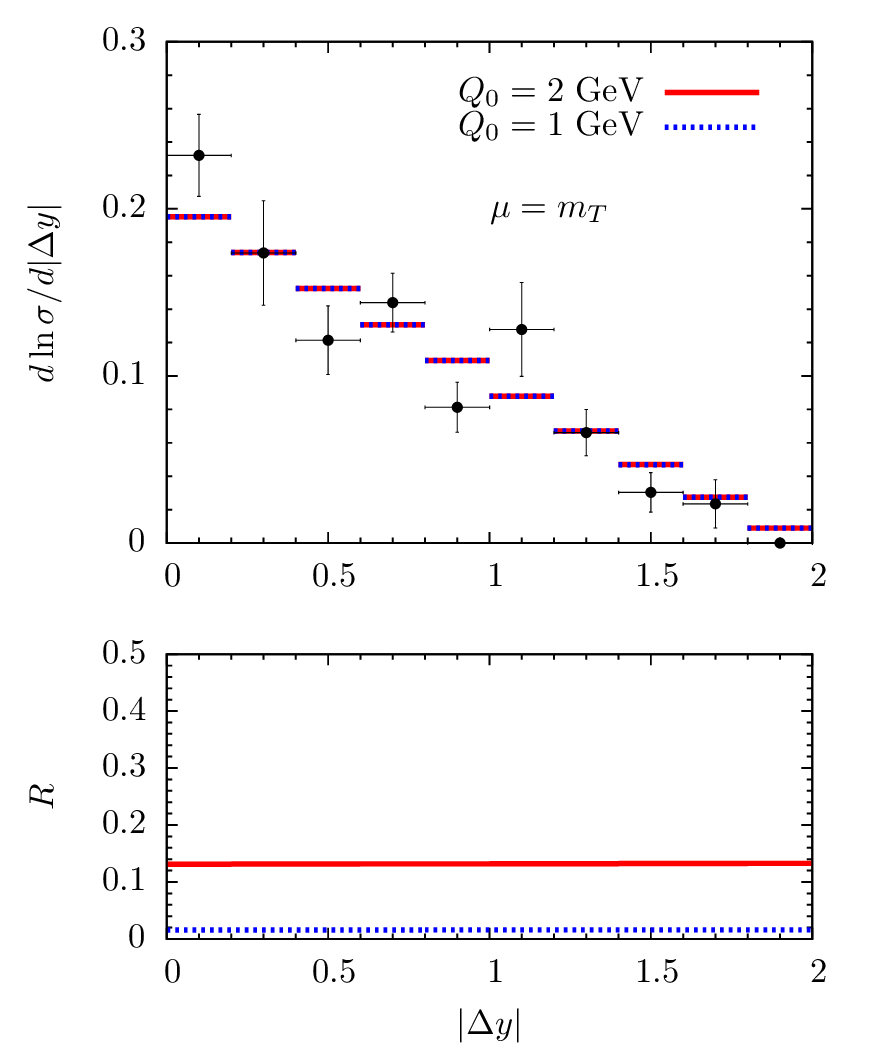}
  \caption{\label{fig:Dy} Same as in figure \ref{fig:pT}, but for the cross section differential in the rapidity difference $\Delta y$ between the two charm quarks.}
\end{figure}

Figure \ref{fig:Dy} shows the dependence of the normalized cross section on the rapidity difference between the charm quarks. The cross section results are stable under variations of the scales and nicely reproduce the shape of the data. The two input scales have a strong impact on the size of the polarized contributions. The relative polarized contribution displays no dependence on the rapidity difference. With $\mu=2m$ the ratio of polarized over unpolarized is 30\% for $Q_0=2\gev$ and around $4\%$ for $Q_0=1\gev$. Changing to $\mu=m_T$ decreases the ratio to about half, $15\%$ for $Q_0=2\gev$ and 2\% for $Q_0=1\gev$. 
\begin{figure}[htb]
  \centering
	\includegraphics[width=0.4\textwidth]{%
      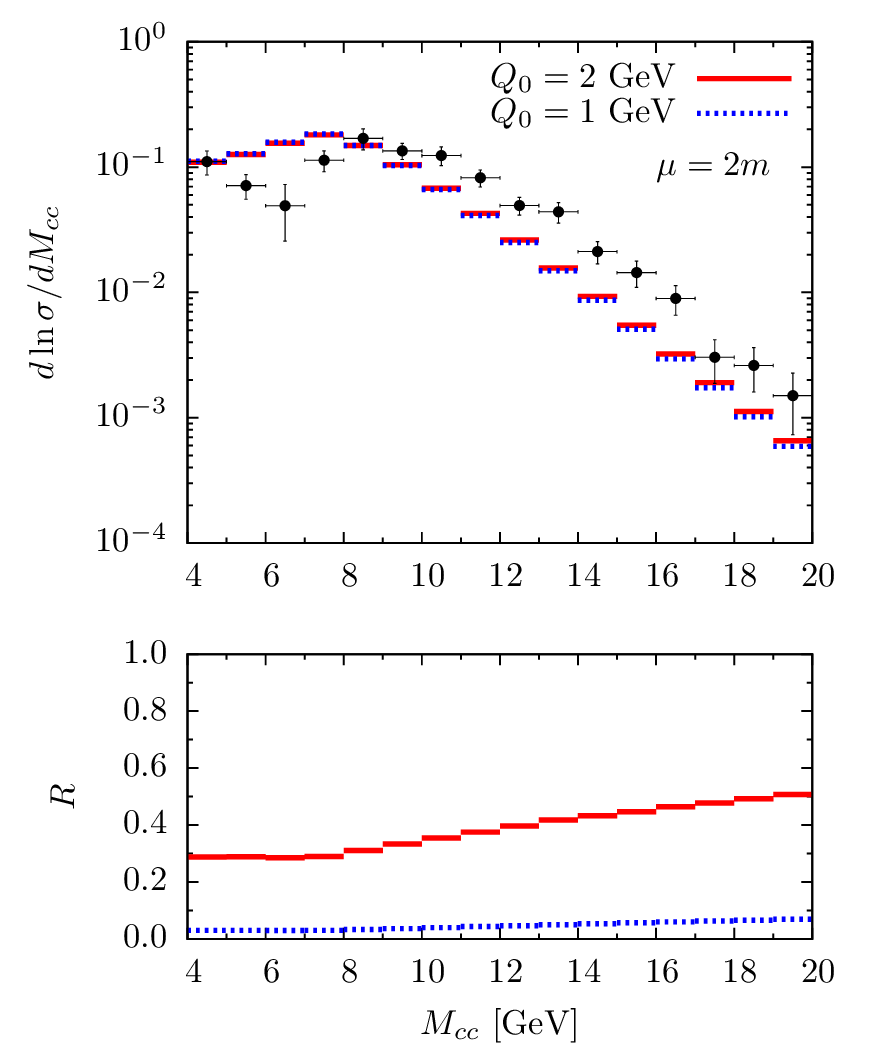}
 	\includegraphics[width=0.4\textwidth]{%
      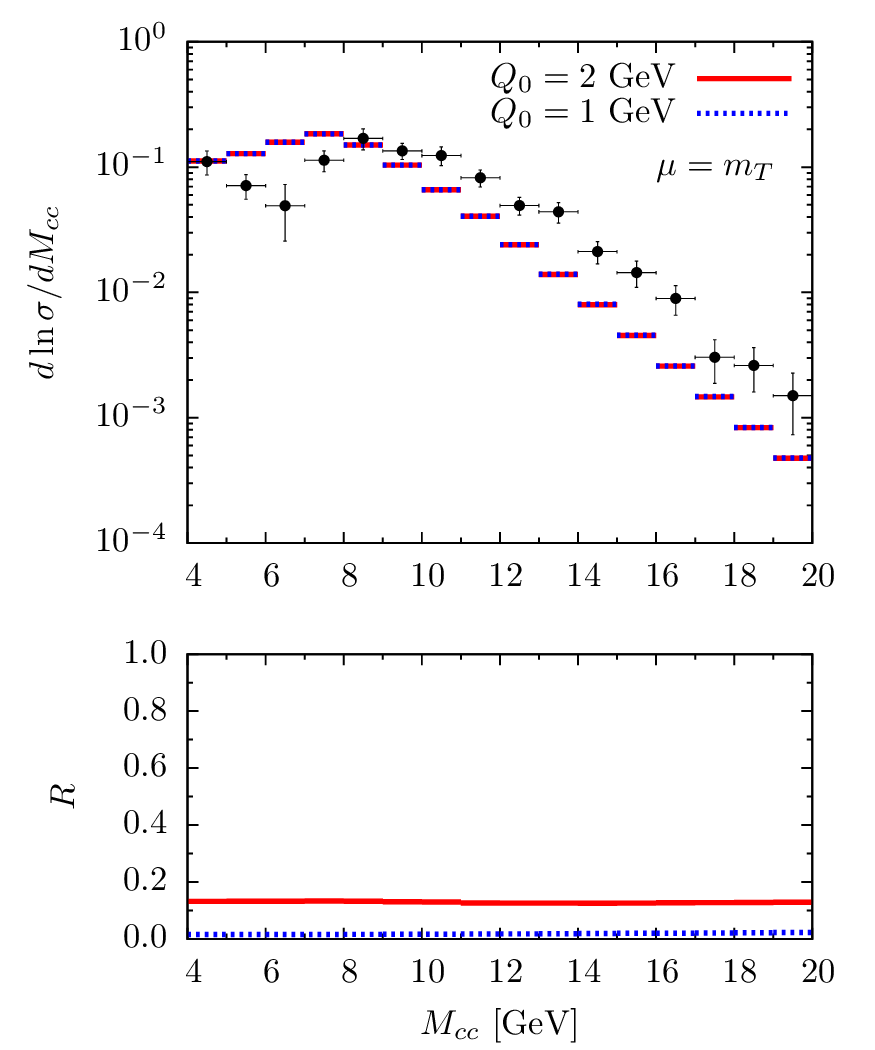}
  \caption{\label{fig:mcc}Same as in figure \ref{fig:pT}, but for the cross section differential in the invariant mass of the two charm quarks $M_{cc}$.}
\end{figure}

The cross section dependence on the invariant mass of the two charm quarks, $M_{cc}$, is shown in figure \ref{fig:mcc}. As in the previous figures, the data is rather well reproduced by the double charm cross section calculation. The polarized contribution has some dependence on $M_{cc}$ with $\mu=2m$ and thus a small impact on the shape of the cross section, but this effect disappears for $\mu=m_T$. With the lower input scale the polarized contribution is a few percent. With the larger $Q_0$ the polarized gluon contribution is $30\%$ of the unpolarized at small $M_{cc}$ and increases up to $50\%$ at large $M_{cc}$ for $\mu=2m$, while $\mu=m_T$ gives a ratio just above $10\%$ in the entire $M_{cc}$ range. 
\begin{figure}[htb]
  \centering
	\includegraphics[width=0.4\textwidth]{%
      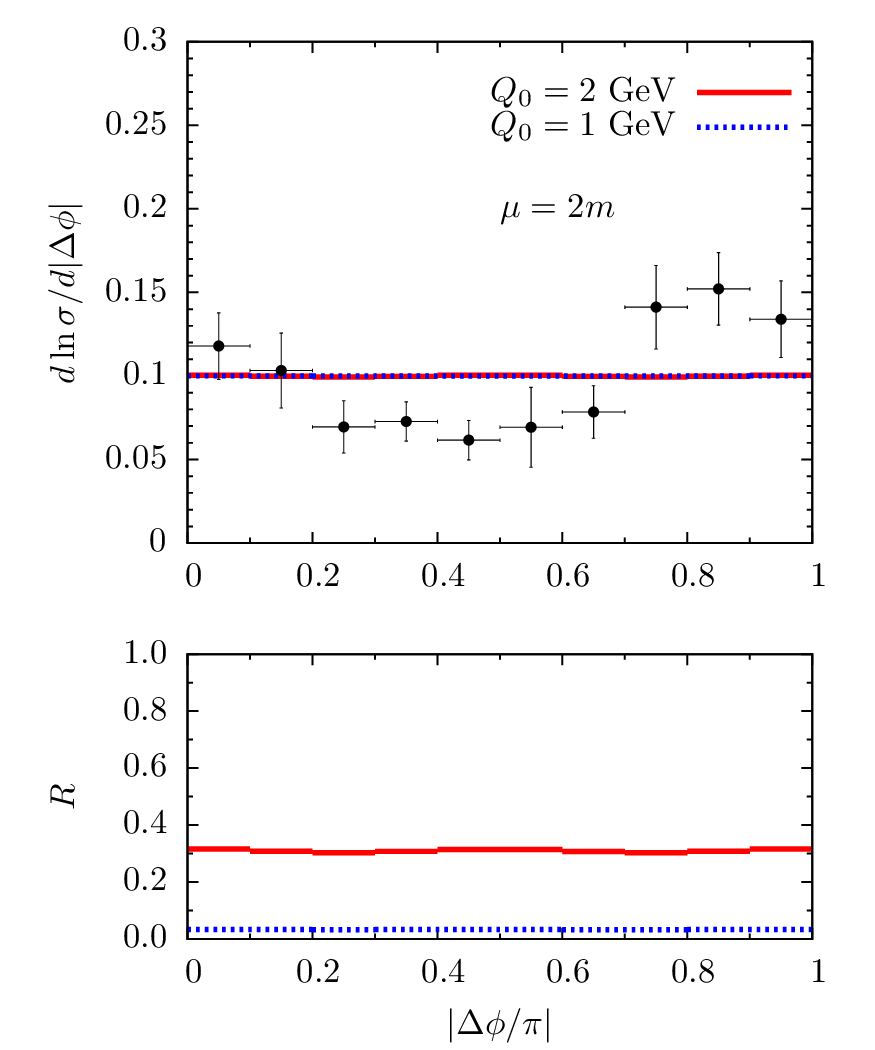}
 	\includegraphics[width=0.4\textwidth]{%
      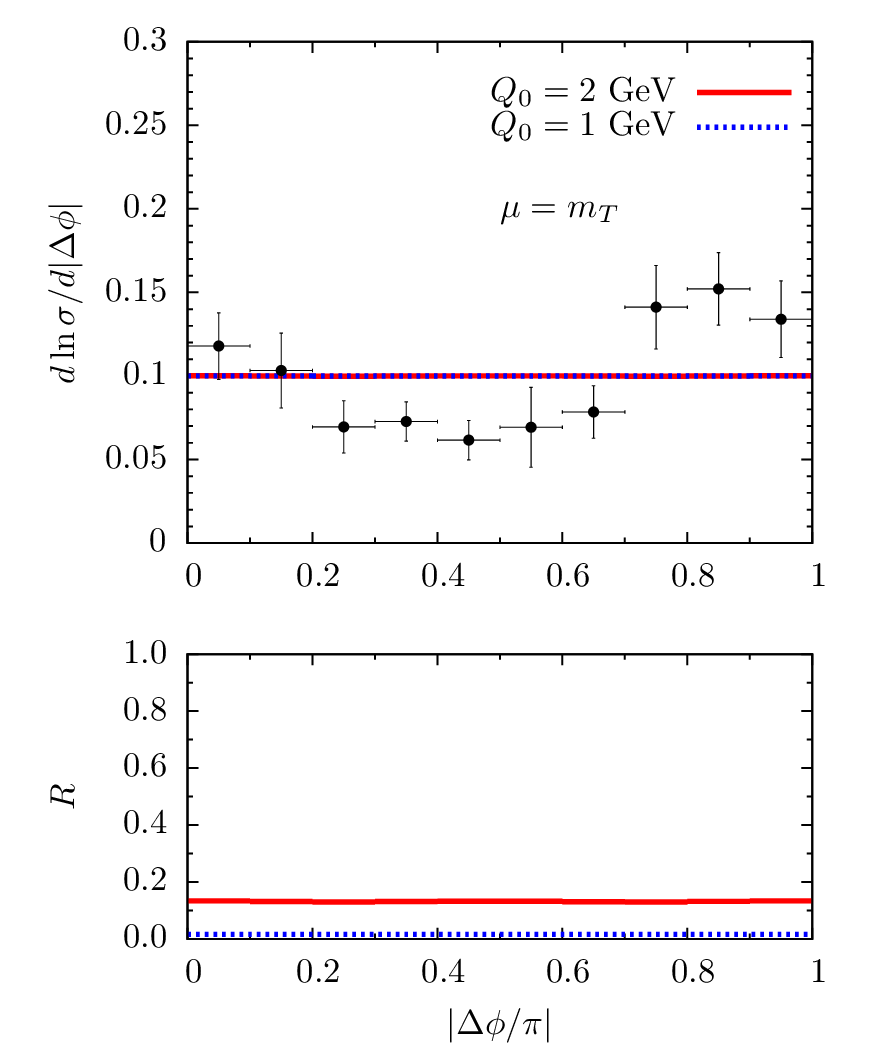}
  \caption{\label{fig:Dphi} Same as in figure \ref{fig:pT}, but for the cross section differential in the azimuthal angle between the two charm quarks.}
\end{figure}	

An intriguing aspect of the $D^0D^0$ results is the azimuthal correlation between the two mesons. This correlation differs from that observed between meson final states with an equal number of charm quarks and anti-quarks, such as $D^0\bar{D}^0$ and $D^\pm D^\mp$ \cite{Aaij:2012dz}, which are dominated by single parton scattering. The angular modulation in $D^0D^0$ resembles that of a $\cos2\Delta\phi$ dependence, which is naturally produced by the DPS cross section involving a mixture of unpolarized and linearly polarized gluons. 
In our LO calculation, the polarized contribution can still be sizable as demonstrated in figure~\ref{fig:Dphi}, but the large contribution originates from the longitudinally polarized gluons and is thus independent of $\Delta \phi$. 
\begin{figure}[htb]
  \centering
	\includegraphics[width=0.4\textwidth]{%
      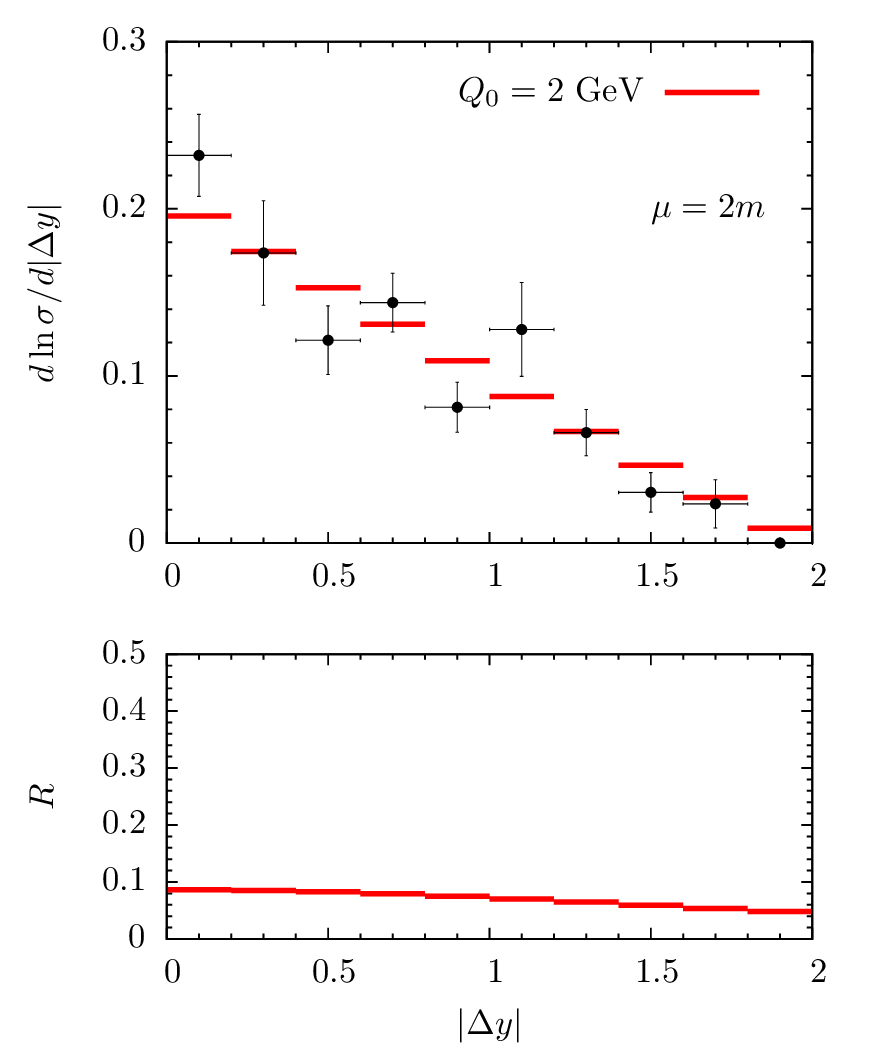}
  \caption{\label{fig:Dy-splitt} Normalized cross section in the splitting scenario vs the difference of rapidity between the two charm quarks at $\mu=2m$ (left). Overlaid are the LHCb $D^0D^0$ data \cite{Aaij:2012dz}. The lower panel shows the relative size of the polarized contribution.}
\end{figure}

Instead of the maximal polarization model for the DPDs, we can use the splitting model - where the ratios of the perturbative splittings of an unpolarized parent parton into two (unpolarized or polarized) partons are the basis of the relations between the different DPDs as explained in \sect{sec:model}. This scenario has a smaller polarization, but different $x_i$ dependences of the unpolarized compared to polarized DPDs - which could show up as shape differences in the rapidity spectrum. Figure~\ref{fig:Dy-splitt} shows the cross section as a function of the rapidity difference, with the polarized DPDs from the splitting model for $Q_0=2\gev$ and $\mu=2m$. The size of the polarized cross section is reduced in the splitting scenario to about 5-10\%, with some dependence on the rapidity difference.

\begin{figure}[htb]
  \centering
	\includegraphics[width=0.4\textwidth]{%
      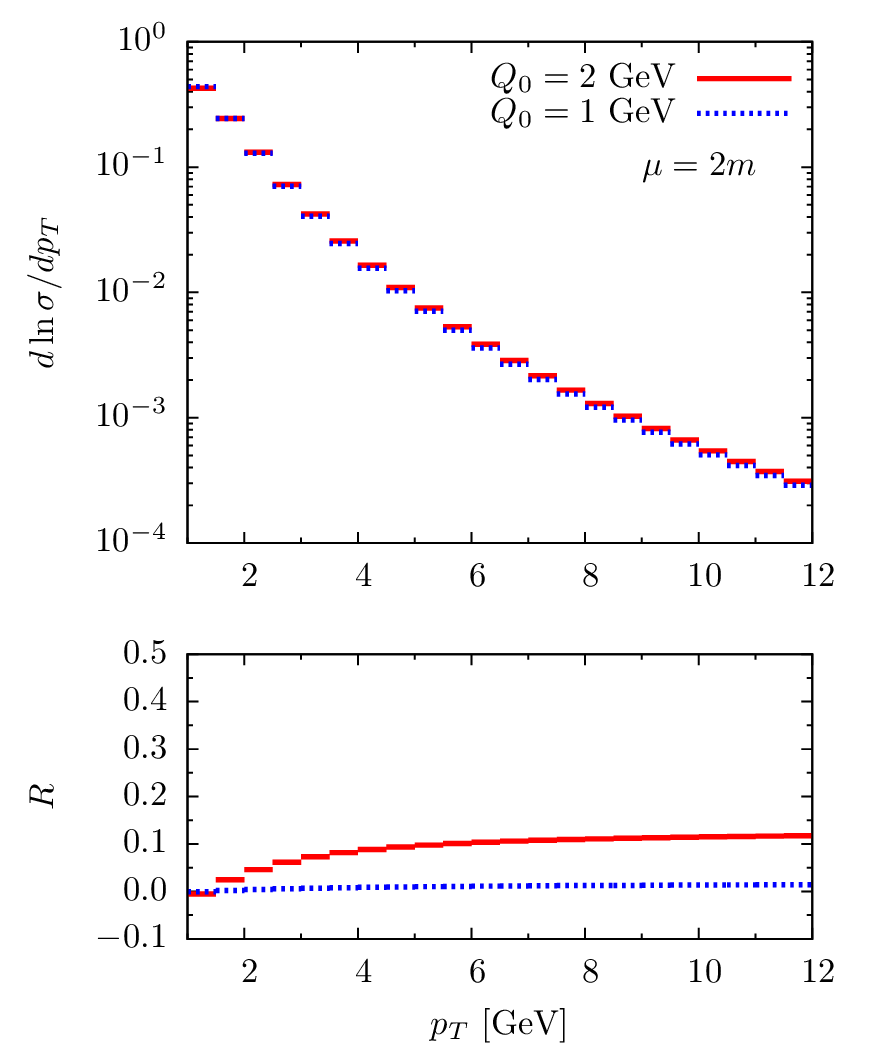}
      \includegraphics[width=0.4\textwidth]{%
      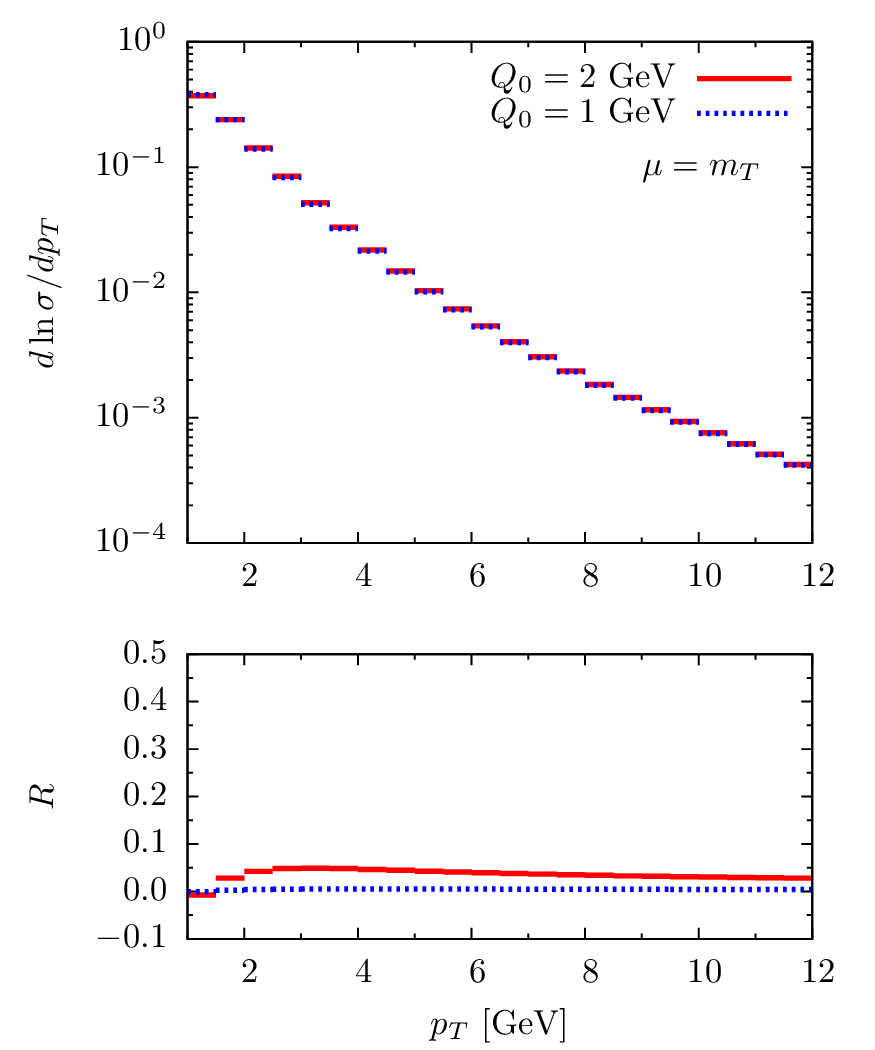}
  \caption{\label{fig:DpT-ext} Normalized cross section vs the transverse momentum of one of the charm quarks in the extended region down to $p_{Ti} = 1\gev$, at $\mu=2m$ (left) and $\mu=m_T$ (right). The lower panels show the relative size of the polarized contribution.}
\end{figure}
Extending the kinematic region to examine in particular the effects of going down towards lower values of $p_T$, figure~\ref{fig:DpT-ext} shows the cross section dependence on $p_T$ in the kinematic range $1\leq p_{Ti} \leq 12 \gev$. This decreases the size of the polarized contribution, which is not surprising since the longitudinally polarized cross section in \eqref{eq:xseclong} has a $(1-2 m/m_{Ti})$ factor for each of the two partonic processes, which decreases when going to smaller $p_{Ti}$.  Although the polarization in this region is rather small, the effect on the shape of the cross section is interesting.
\begin{figure}[htb]
  \centering
	\includegraphics[width=0.4\textwidth]{%
      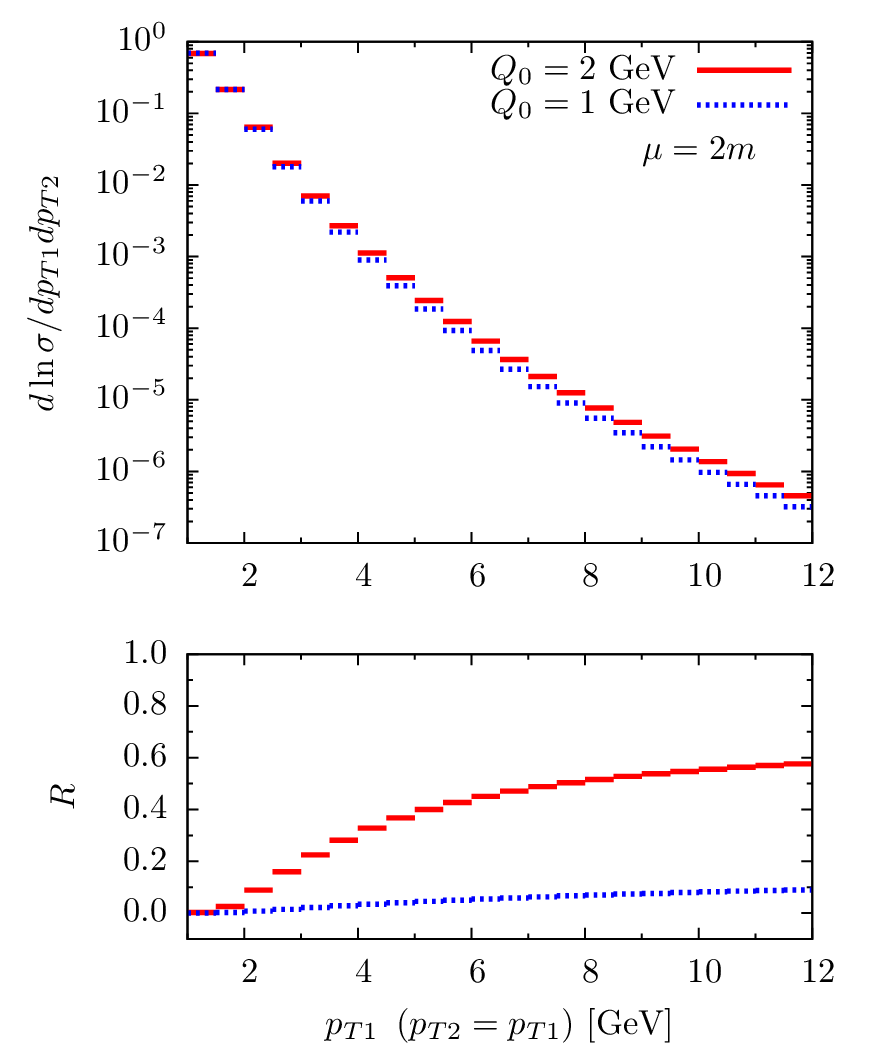}
      \includegraphics[width=0.4\textwidth]{%
      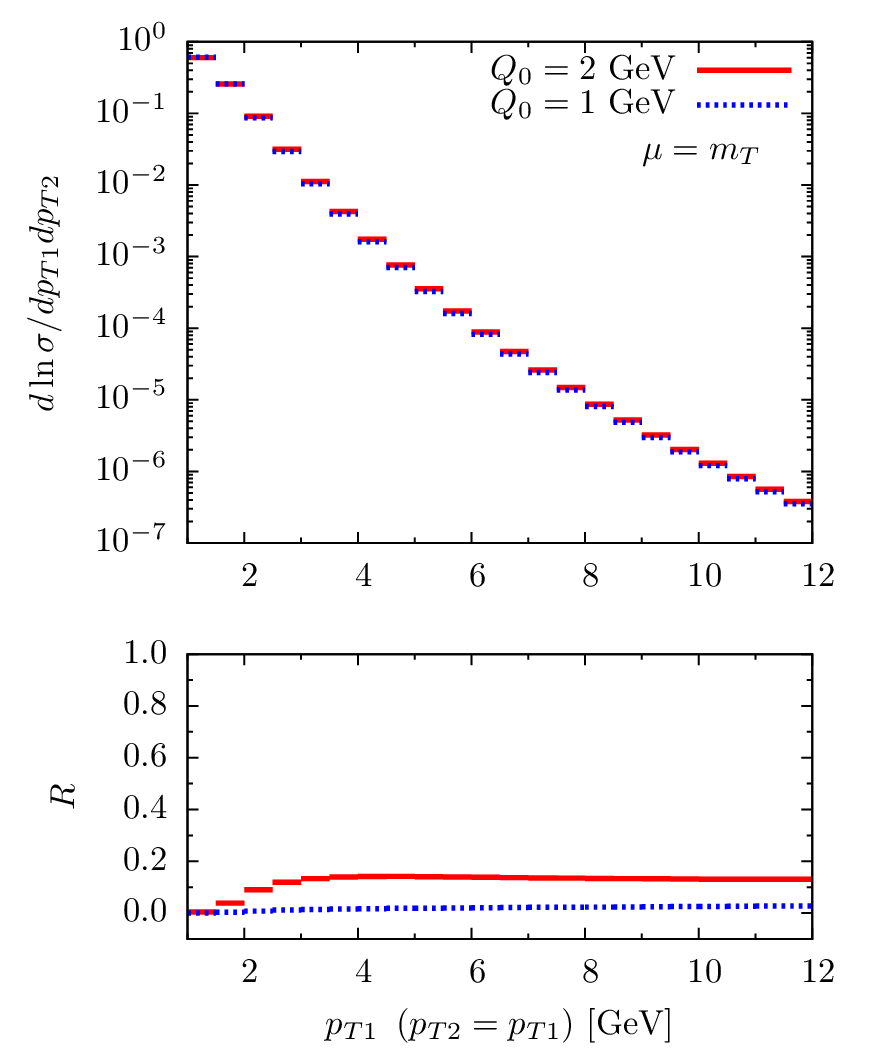}
  \caption{\label{fig:DDpT} Normalized double differential cross section vs the transverse momentum of the two charm quarks in the extended region down to $p_{Ti} = 1\gev$, at $\mu=2m$ (left) and $\mu=m_T$ (right). The lower panels show the relative size of the polarized contribution. }
\end{figure}
In figure~\ref{fig:DDpT} we show the double differential cross section, in $p_{T1}$ and $p_{T2}$. We see a strong $p_{Ti}$ dependence of the polarized contribution in combination with a large absolute size, which starts at $0\%$ for $p_{Ti}=1\gev$ and goes up to $60\%$ of the unpolarized for $p_{Ti}$ approaching 12 GeV, with $\mu=2m$ and $Q_0=2\gev$. The results with $\mu=m_T$ have less polarization, with a maximal ratio reduced to about 10\%. Some of the $p_{Ti}$ dependence remains but most of it is at $p_{Ti}$ values below $3\gev$. A precise measurement of this double differential cross section could be able to distinguish some of the different scenarios, and either see first indications of or set first limits on the longitudinally polarized gluons in DPS. In particular if it is possible to extend the measured region down to lower transverse momenta. 
\begin{figure}[htb]
  \centering
	\includegraphics[width=0.4\textwidth]{%
      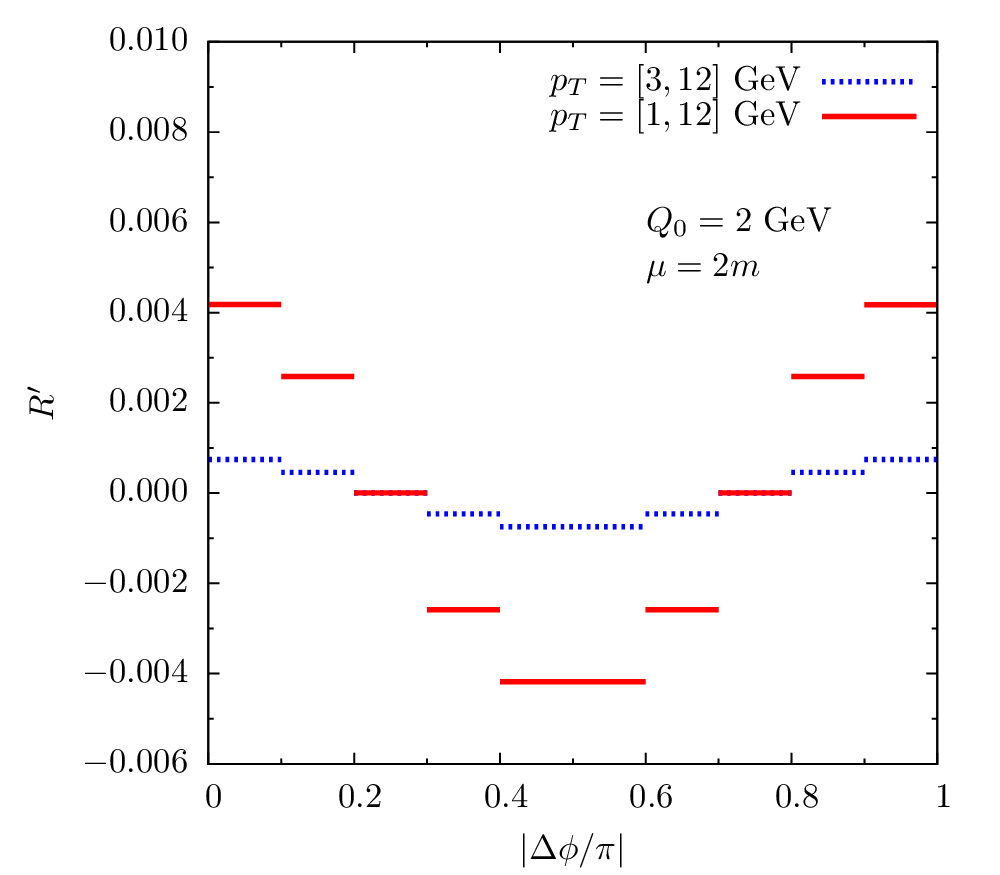}
  \caption{\label{fig:Dphi-ext} Normalized cross section vs the azimuthal angle between the two charm quarks, at $\mu=2m$, with kinematic region extended down to $p_{Ti} = 1\gev$. }
\end{figure}
Investigating the dependence of the relative size of the mixed unpolarized - linearly polarized gluons on the lower limit of transverse momenta we expect a rather large increase in the relative size (compared to the cross section contribution without azimuthal dependence). This is also visible in figure~\ref{fig:Dphi-ext}, where the relative size of the mixed contribution is increased by almost an order of magnitude when extending the kinematic region down to $p_{Ti}=1\gev$. The amplitude is still small compared to the angular modulation in the data, but it is another indication that allowing for a non-zero transverse momentum of the initial gluons, through for example NLO correction, could lead to significant enhancements.

\subsection{Predictions at $\sqrt{s}=13$ TeV}
\begin{figure}[htb]
  \centering
	\includegraphics[width=0.4\textwidth]{%
      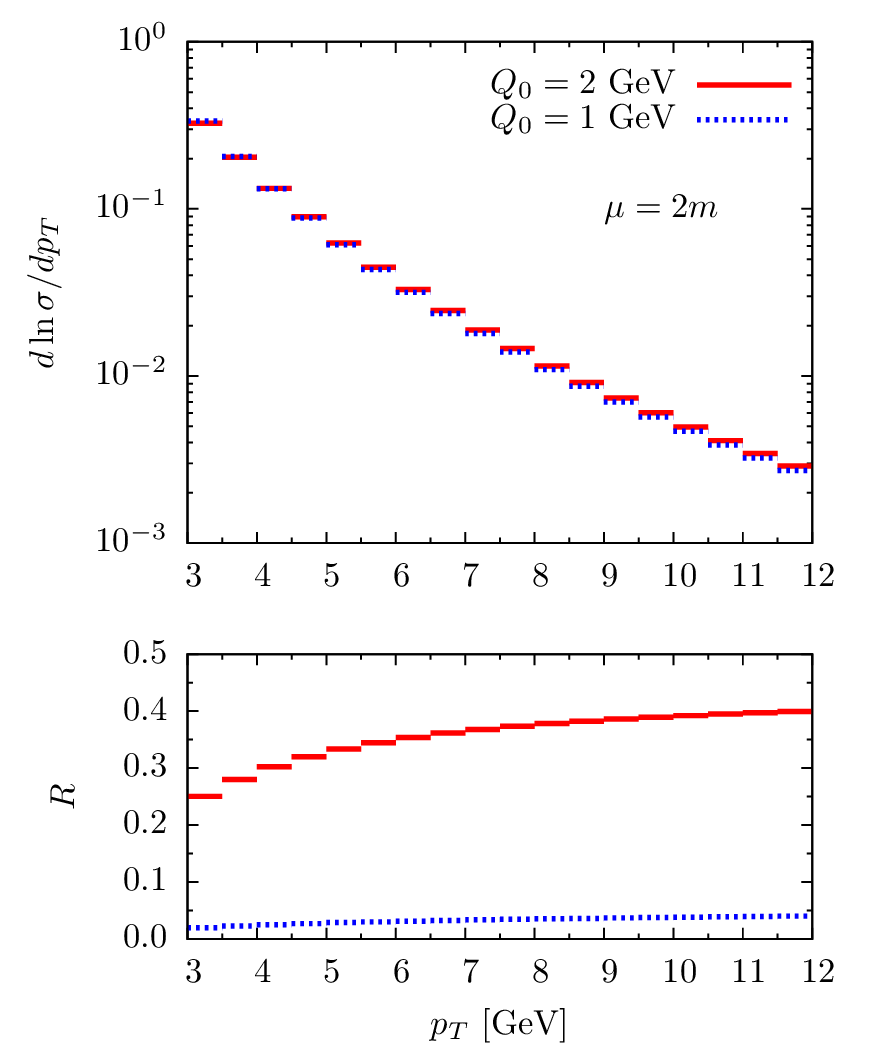}
 	\includegraphics[width=0.4\textwidth]{%
      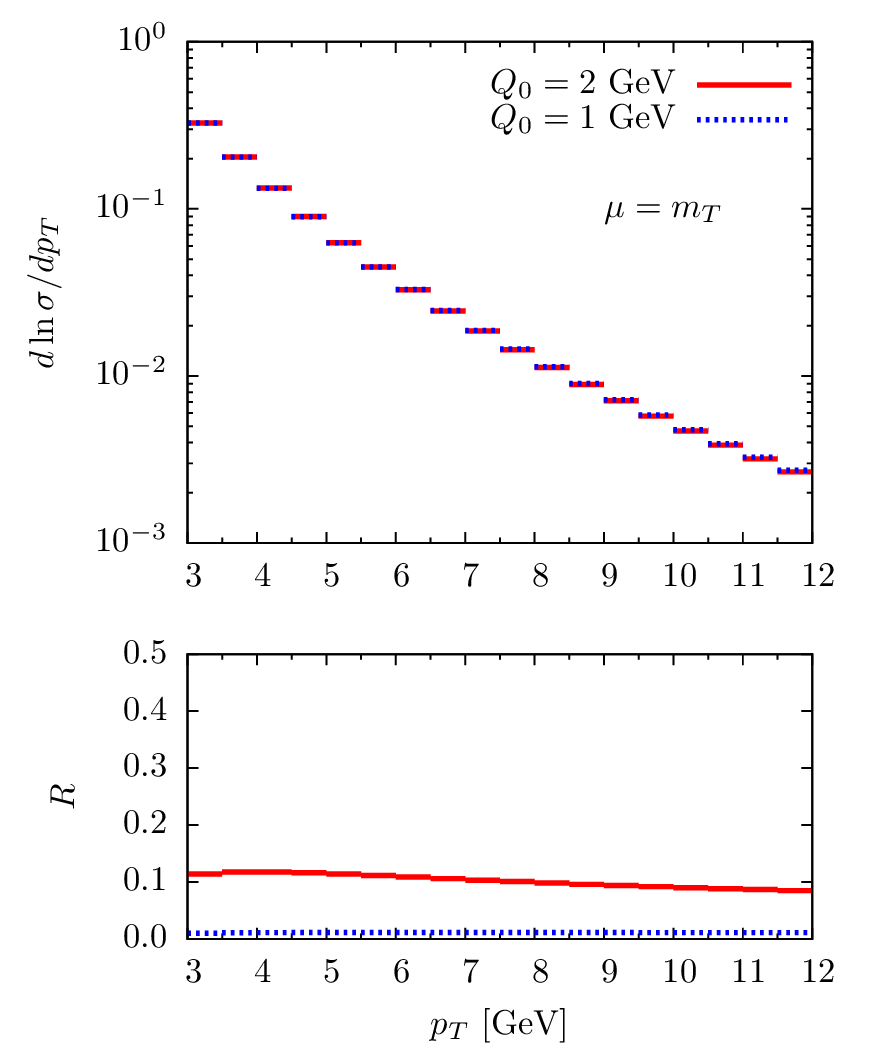}
      	\includegraphics[width=0.4\textwidth]{%
      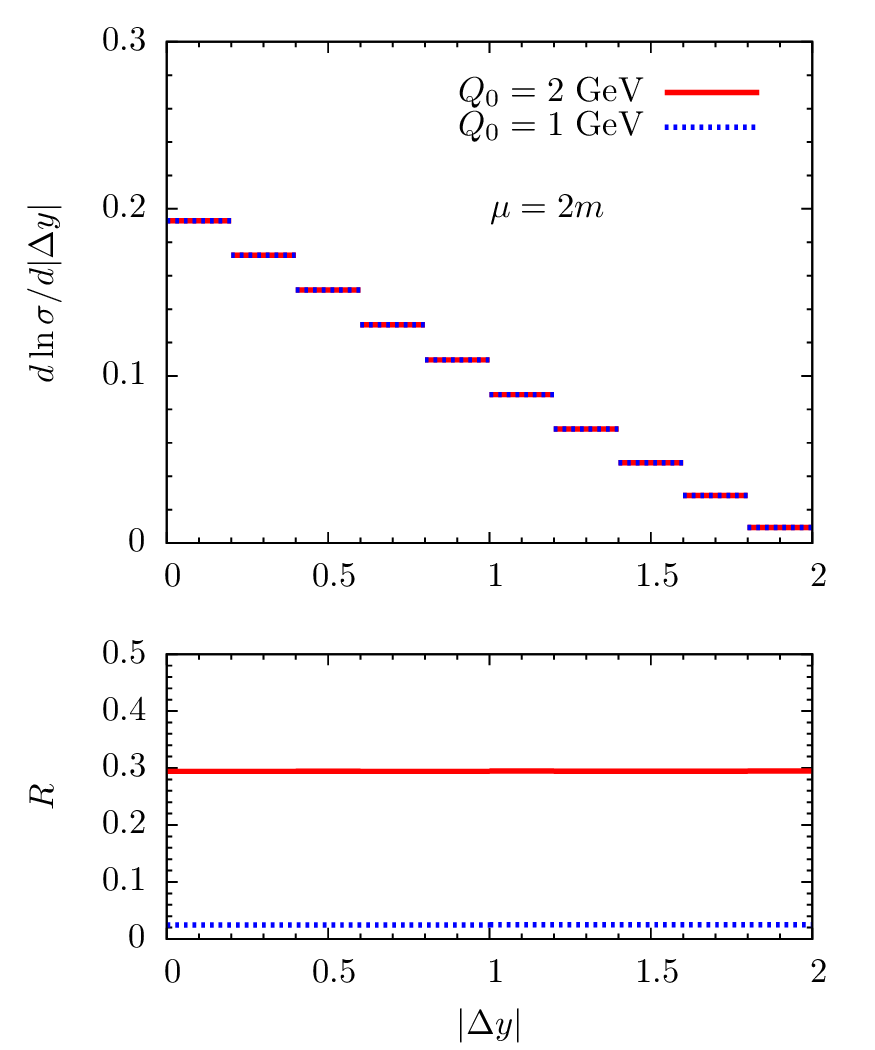}
 	\includegraphics[width=0.4\textwidth]{%
      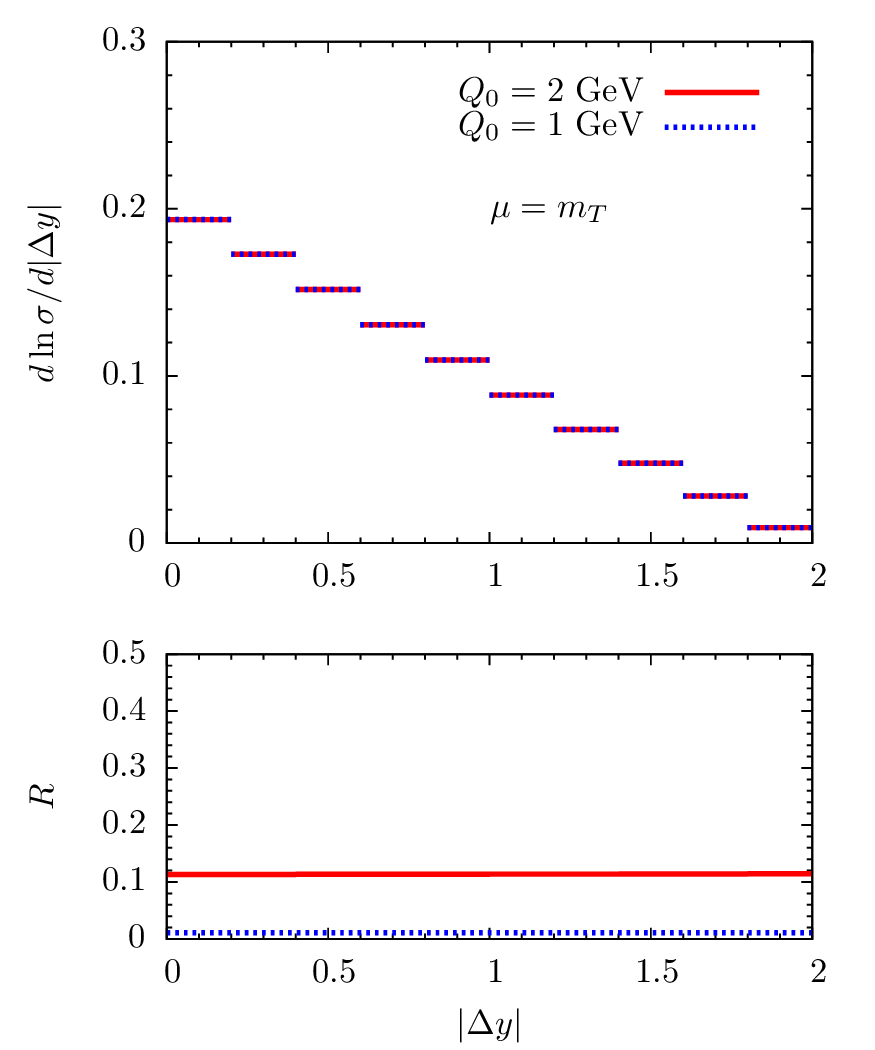}
  \caption{\label{fig:pT13} Collisions at $\sqrt{s}=13$~TeV. Top: normalized cross section vs the transverse momentum of one of the charm quarks with $\mu=2m$ (left) and $\mu=m_T$ (right). The lower panels show the relative size $R$ of the polarized contribution. Bottom: normalized cross section vs the rapidity difference $\Delta y$ between the two charm quarks with $\mu=2m$ (left) and $\mu=m_T$ (right) and relative size $R$ of the polarized contribution. }
\end{figure}

\begin{figure}[htb]
  \centering
	\includegraphics[width=0.4\textwidth]{%
      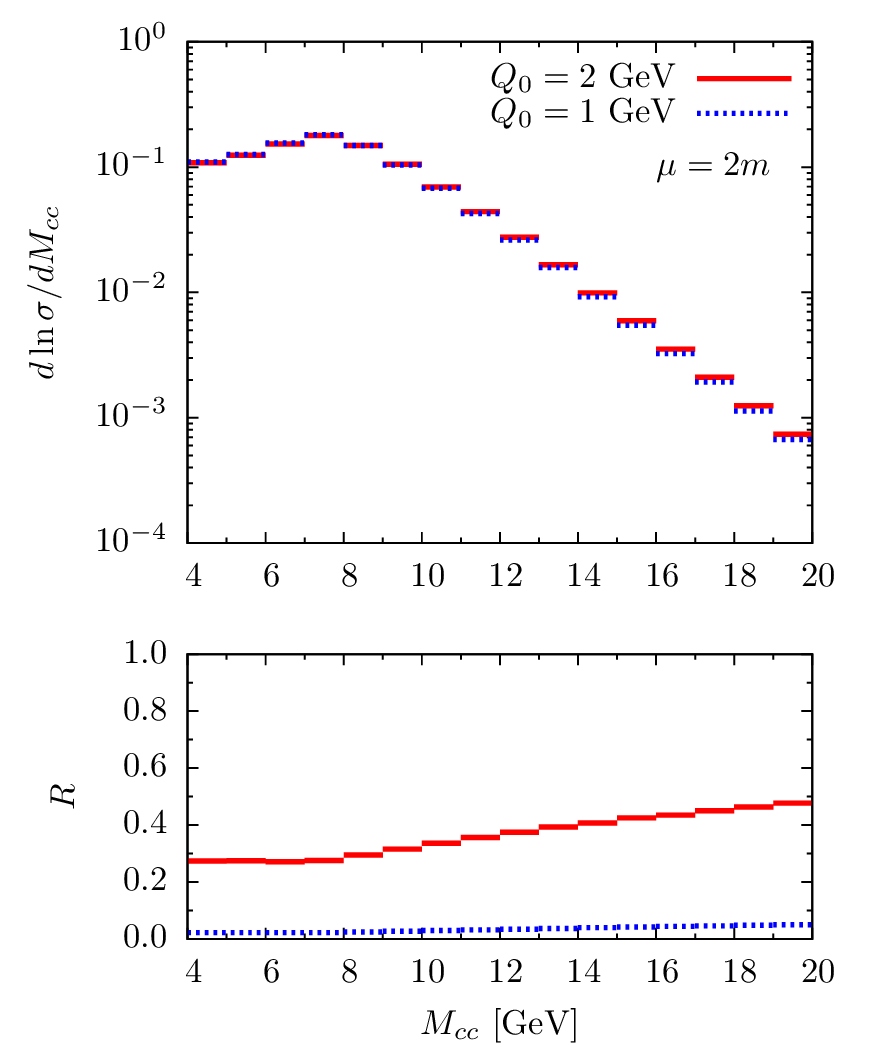}
 	\includegraphics[width=0.4\textwidth]{%
      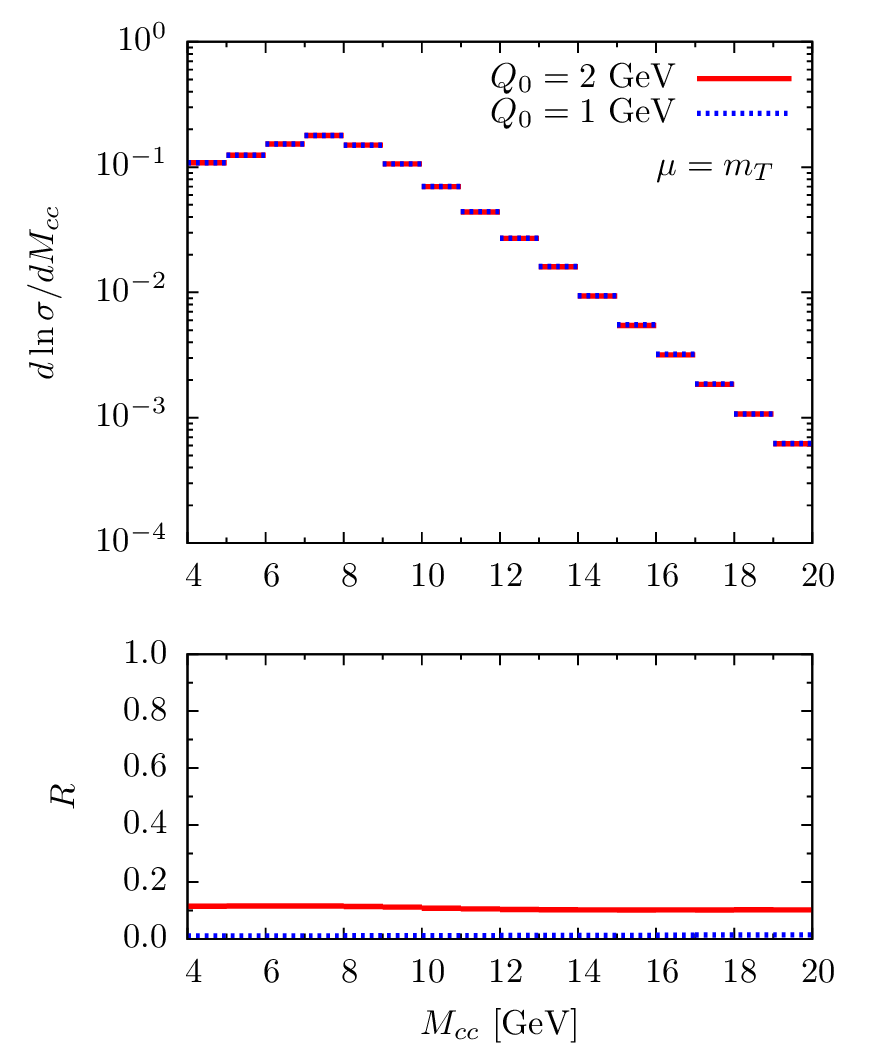}
      	\includegraphics[width=0.4\textwidth]{%
      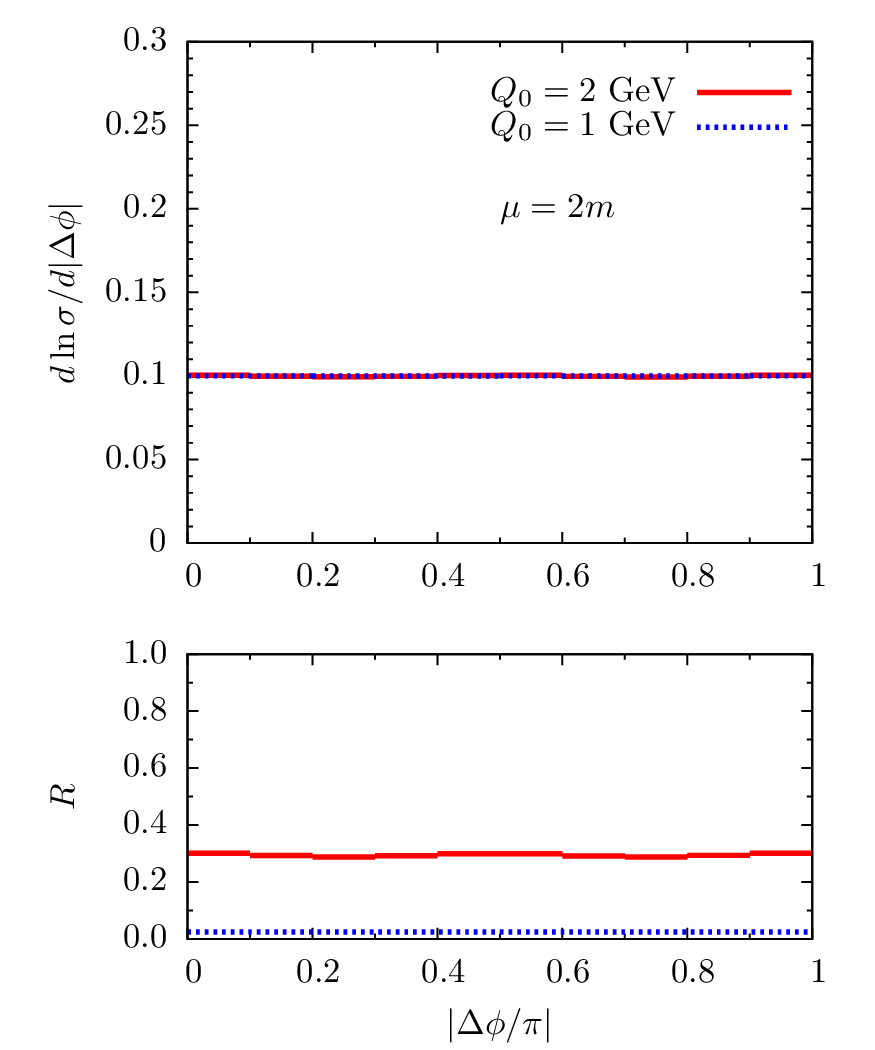}
 	\includegraphics[width=0.4\textwidth]{%
      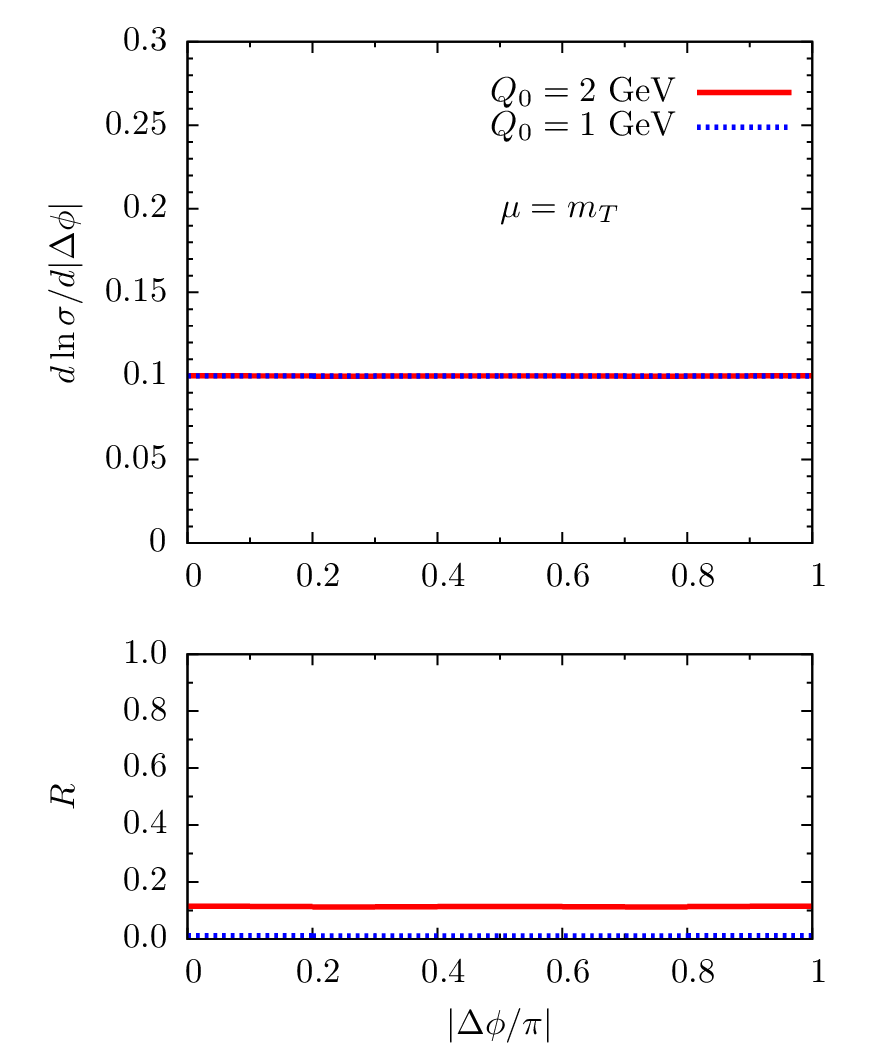}
  \caption{\label{fig:mcc13} Collisions at $\sqrt{s}=13$~TeV. Top: normalized cross section vs the invariant mass of the two charm quarks with $\mu=2m$ (left) and $\mu=m_T$ (right). The lower panels show the relative size $R$ of the polarized contribution. Bottom: normalized cross section vs the azimuthal angle between the two charm quarks $\Delta \phi$ between the two charm quarks with $\mu=2m$ (left) and $\mu=m_T$ (right) and relative size $R$ of the polarized contribution.}
\end{figure}

In this section we show predictions at a hadronic center of mass energy of $13$~TeV. The results are generally very similar to those at $\sqrt{s}=7$~TeV, and we will therefore keep the discussion rather brief. Figure \ref{fig:pT13} (first row) displays the normalized cross section as a function of the $p_T$ of one of the two quarks. The change in CM energy as compared to figure \ref{fig:pT} flattens the cross section slightly and leads to a small decrease of the polarization. The second row of figure \ref{fig:pT13} shows the cross section results as a function of the rapidity difference $\Delta y$. The change in CM energy has no visible impact on the shape of the cross section, and only leads to a small decrease of the polarized contribution.
This small decrease of the polarization as well as the small, if any, changes to the shape of the cross section is observed also for the dependence on $M_{cc}$ and $\Delta \phi$, as demonstrated by figure \ref{fig:mcc13}.

\begin{figure}[htb]
  \centering
	\includegraphics[width=0.4\textwidth]{%
      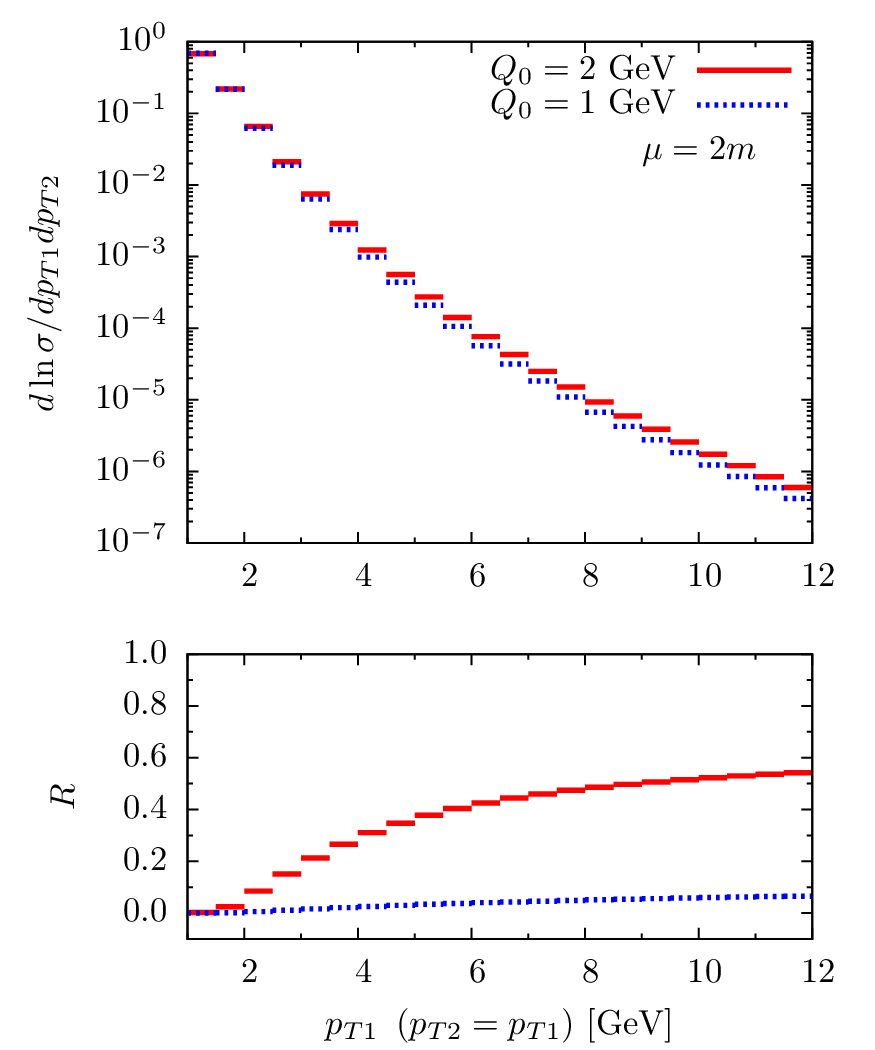}
      \includegraphics[width=0.4\textwidth]{%
      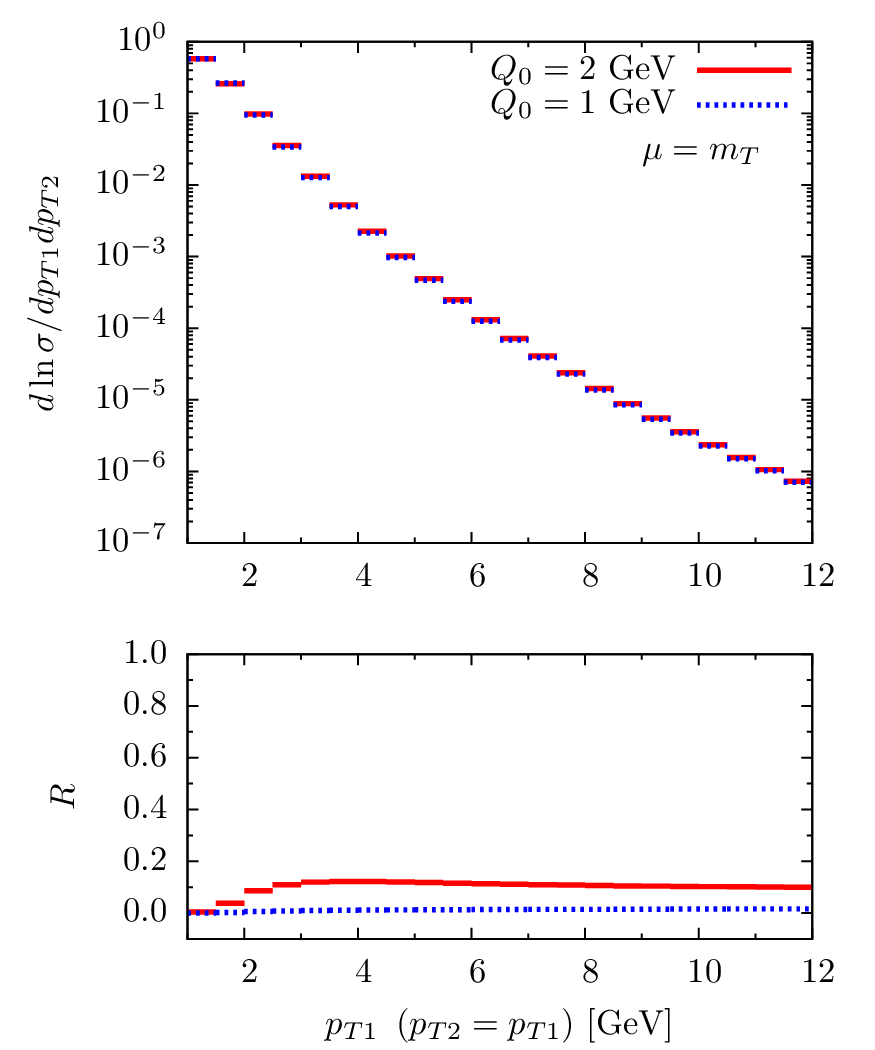}
  \caption{\label{fig:DDpT13} Collisions at $\sqrt{s}=13$~TeV. Normalized double differential cross section vs the $p_{T}$ of the two charm quarks, with $\mu=2m$ (left) and $\mu=m_T$ (right), in the extended region down to $p_{Ti} = 1\gev$.}
\end{figure}

In the extended $p_{Ti}$ region for the double differential cross section, the results have large polarization with a strong dependence on the transverse momentum at $\mu=2m$ and $Q_0=2\gev$, as shown in figure \ref{fig:DDpT13}. The contribution of the polarization decreases, as does the shape dependence, when going to $\mu=m_T$.

\section{Conclusions}
\label{sec:concl}
We have investigated the effects of polarization in the double open-charm cross section, when the two charm quarks are produced in the kinematic region probed in the $D^0D^0$ measurement by the LHCb Collaboration \cite{Aaij:2012dz}. Polarization can give sizable effects on the magnitude of the cross section, reaching up above $50\%$ of the unpolarized contribution in certain kinematic regions. The size strongly depends on the choices made when modeling the polarized double gluon distributions and on the large uncertainties for the single gluon distributions at the relevant low scales and small momentum fractions. We have presented the results obtained with $\sqrt{s}$ equal to both $7$ and $13$ TeV. The change of energy scale only has minor impact on the shape of the DPS cross section results as well as the relative size of the polarized contributions.

The shape of the polarized contributions to the cross section are in most variables quite similar to the unpolarized results. In these cases it is difficult to disentangle the polarized contribution from other contributions in the DPS cross section, such as a single parton splitting and color interference contributions. We therefore identify variables and kinematic regions where the polarization does introduce some shape dependence. The most prominent shape dependence is found for the cross section double differential in the $p_T$ of the two charm quarks, where the polarized contribution can vary with $p_T$ from 0 up to 60\% of the unpolarized.

We compare the results of our calculation with the measurement of $D^0D^0$ mesons by the LHCb \cite{Aaij:2012dz}. For most distributions, the leading order calculation reproduces the experimental data rather well. The data cannot discriminate between the different models for the polarized DPDs as the polarization does not introduce any strong shape changes. The exception is the dependence on the azimuthal angle between the two mesons, which exhibits an approximate $\cos2\Delta\phi$ modulation. Polarized double parton scattering naturally produces such a modulation in the combination of linearly polarized and unpolarized gluons. However, the leading order DPS cross section for this term is too small to reproduce the modulation in the data. It is possible, however, that the size of this term changes drastically when including higher orders. Higher order effects for the process are expected to be large \cite{vanHameren:2014ava}. Such large NLO corrections, in combination with an expectation that the higher order corrections will lift the strong suppression of the mixed (unpolarized - linearly polarized) contribution present at tree level, can lead to a significant enhancement of the amplitude of the azimuthal modulation. Unfortunately, the theoretical formalism for the description of DPS needs to be further developed to reach a state where higher order effects can be systematically included.

In the double differential cross section, looking at the $p_{T}$ of both of the charm quarks, the longitudinal polarization can have a larger impact on both the size and shape. Measurements of this double differential cross section could therefore give first experimental indications of, or limits on, the effects of polarization in double parton scattering.

We have used the homogeneous double DGLAP evolution equations, which do not include any single part splitting term. The effect of the single parton splitting on the unpolarized DPS cross section was studied in \cite{Gaunt:2014rua}. Including it also for the polarized terms of the DPS cross section could further enhance the effects of polarization. In addition, we have employed an ansatz which splits the unpolarized gluon DPD in two single gluon PDFs and a factor depending only on the transverse distance between the two partons. This approach, common in DPS studies, is useful as a first approximation of the gluon DPD, but neglects several effects. These include correlations between kinematical variables and the color of the two gluons. The cross section ratios which we present, are likely to be more stable to such corrections than the absolute size of the cross section. However, further phenomenological as well as experimental studies are required to better constrain these effects.

\section*{Acknowledgements}
Many thanks to Daniel Boer, Markus Diehl, Robert Fleischer, Francesco Hautmann and Hannes Jung for useful discussions and Vanya Belyaev for supplying us with the LHCb data points. We are in debt to Jonathan Gaunt for providing us with his original evolution code. We acknowledge
financial support from the European Community under the ``Ideas'' program
QWORK (contract 320389). M.G.E.~is supported by the ``Stichting voor Fundamenteel Onderzoek der Materie'' (FOM), which is financially supported by the ``Nederlandse Organisatie voor Wetenschappelijk Onderzoek'' (NWO). C.P. acknowledges support by the "Fonds Wetenschappelijk Onderzoek - Vlaanderen" (FWO) through a postdoctoral Pegasus Marie Curie Fellowship. Figures were made using JaxoDraw \cite{Binosi:2003yf}.

\appendix

\bibliography{../../LatexStuff/KasBib}

\end{document}